\documentclass[useAMS,usenatbib]{mn2e}
\usepackage[colorlinks,urlcolor=blue,citecolor=blue,linkcolor=blue]{hyperref}
\usepackage[pdftex]{graphicx}
\usepackage{epstopdf,color,verbatim}
\usepackage{threeparttable}
\usepackage{psfrag}

\usepackage[]{txfonts}
\usepackage{textcomp}
\def\simless{\mathbin{\lower 3pt\hbox
{$\rlap{\raise 5pt\hbox{$\char'074$}}\mathchar"7218$}}}   
\def\simmore{\mathbin{\lower 3pt\hbox
{$\rlap{\raise 5pt\hbox{$\char'076$}}\mathchar"7218$}}}   

\newcommand{\eqb}{\begin{eqnarray}}
\newcommand{\eqe}{\end{eqnarray}}
\newcommand{\be}{\begin{eqnarray}}
\newcommand{\ee}{\end{eqnarray}}
\newcommand{\bi}{\begin{itemize}}
\newcommand{\ei}{\end{itemize}}
\newcommand{\mel}{m_{\rm e}}
\newcommand{\mpr}{m_{\rm p}}

\newcommand{\Bp}{B^{\prime \prime}_{\rm p}}
\newcommand{\dpr}{\prime \prime}

\newcommand{\sth}{\sigma_{\rm T}}

\newcommand{\gstar}{\gamma_\star}

\newcommand{\wf}{w^{\prime \prime}_{\rm f}}
\newcommand{\gmin}{\gamma_{\min}}
\newcommand{\gmax}{\gamma_{\max}}
\newcommand{\gbr}{\gamma_{\rm br}}

\newcommand{\thobs}{{\theta}_{\rm obs}}

\newcommand{\dpl}{\delta_{\rm p}}
\newcommand{\Gp}{\Gamma_{\rm p}}
\newcommand{\Gj}{\Gamma_{\rm j}}
\newcommand{\Gco}{\Gamma_{\rm co}}
\newcommand{\bco}{\beta_{\rm co}}
\newcommand{\vg}{\beta_{\rm g}}
\newcommand{\vacc}{\beta_{\rm acc}}
\newcommand{\vexp}{\beta_{\rm exp}}
\newcommand{\vexpo}{\beta_{\rm exp,0}}

\newcommand{\ksone}{k_{\rm s, I}}
\newcommand{\kstwo}{k_{\rm s, II}}

\newcommand{\Dtdb}{\Delta {\tau}_{1/2}}

\usepackage{graphicx}
\usepackage{epsfig} 
\usepackage{epstopdf}

\title[Blazar flares from plasmoids]
{Blazar flares powered by plasmoids in relativistic reconnection}

\author[Petropoulou, Giannios, \& Sironi]
{Maria Petropoulou$^{1}$\thanks{Einstein Post Doctoral Fellow}\thanks{E-mail: mpetropo@purdue.edu},
 Dimitrios Giannios$^{1}$ and Lorenzo Sironi$^2$\\
$^{1}$Department of Physics and Astronomy, Purdue University, 525 Northwestern
Avenue, West Lafayette, IN 47907, USA\\
$^2$ Department of Astronomy, Columbia University, 550 W 120th St, New York, NY 10027, USA
}

\begin{document}
\date{Received / Accepted}
\pagerange{\pageref{firstpage}--\pageref{lastpage}} \pubyear{2016}

\maketitle

\label{firstpage}

\begin{abstract}
Powerful flares from blazars with short ($\sim$min) variability timescales are challenging for current models of blazar emission. Here, we present a physically motivated {\sl ab initio} model for blazar flares based on the results of recent particle-in-cell (PIC) simulations of relativistic magnetic reconnection.  
PIC simulations demonstrate that quasi-spherical plasmoids filled with high-energy particles and magnetic fields are a self-consistent by-product of the reconnection process. By coupling our PIC-based results (i.e., plasmoid growth, acceleration profile, particle and magnetic content) with a kinetic equation for the evolution of the electron distribution function we demonstrate  that relativistic reconnection in blazar jets can produce powerful flares whose temporal and spectral properties are consistent with the observations.  In particular, our model predicts correlated synchrotron and synchrotron self-Compton flares of duration of several hours--days powered by the largest and slowest moving plasmoids that form in the reconnection layer. Smaller and faster plasmoids produce flares of sub-hour duration with higher peak luminosities than those powered by the largest plasmoids. Yet, the observed fluence in both types of flares is similar.  Multiple flares with a range of flux-doubling timescales (minutes to 
several hours) observed over a longer period of flaring activity (days or longer) may be used as a probe of the reconnection layer's orientation and the jet's magnetization. Our model shows that blazar flares are naturally expected as a result of magnetic reconnection in a magnetically-dominated jet. 
\end{abstract} 


\section{Introduction}
Blazars are a small subclass of Active Galactic Nuclei (AGN),  yet they attract an ever growing interest as they are found in increasingly large numbers by surveys at microwaves and $\gamma$-ray energies \citep[e.g.][]{giommiWMAP_09, abdo_10, giommiPlanck12, ackermann_15}. Blazars also represent  the most abundant population of extragalactic sources at TeV energies\footnote{http://tevcat.uchicago.edu/} \citep[e.g.][]{holder_ICRC13, deNaurois_ICRC15}.
The extreme observational properties of blazars, such as continuum emission over the entire electromagnetic spectrum, rapid and large-amplitude variability, make them to stand out among other AGN.  The blazar broadband emission, from radio up to very high energy (VHE) $\gamma$-rays ($>$100~GeV),
is believed to originate from a relativistic jet that is nearly aligned with the observer's line of sight and emerges from the central supermassive black hole 
\citep{blandfordrees78, urry_padovani95}. The blazar spectral energy distribution (SED) is also very distinctive due to its double-hump appearance. A typical blazar SED is composed of two broad components: a low-energy and a high-energy component extending, respectively,  from radio to UV/X-rays and from X-rays to $\gamma$-rays.

Blobs, or quasi-spherical emission regions containing relativistic particles and magnetic fields, have been often invoked to explain the broadband variable emission of blazar jets\footnote{There are also certain models  that aim at explaining the low-energy \citep{marscher_gear85,marscher_travis96} or/and the high-energy \citep[e.g.][]{maraschi_92, reynoso_11, potter_12, reynoso_12} blazar emission in terms of a jet model where the  particle  population dynamically evolves along the jet from its base to $\sim$pc scales. } \citep[e.g.][]{bloom_marscher96, mastkirk_97, kirk_98, chiaberge_99, boettcher_chiang02}. Yet, their physical origin is still not understood. {Here, we present a physically motivated model for the  ``emitting blobs'' in  a Poynting-flux dominated blazar jet \citep[for details see][]{giannios_09, giannios_13}, in the context of relativistic magnetic reconnection; the latter refers to the regime where the magnetic energy per particle exceeds its rest mass energy, or the plasma magnetization 
$\sigma$ exceeds unity\footnote{The plasma magnetization is defined as 
$\sigma=B_0^{\prime 2}/4 \pi \rho^\prime c^2$, where $B_0^{\prime}$ and $\rho^\prime$ are the magnetic field and mass density of the plasma outside the reconnection layer.  These are measured in the rest frame of the jet fluid.}.} MHD instabilities of a Poynting-flux dominated flow lead to the formation of current sheets where magnetic reconnection is triggered \citep[e.g.][]{eichler_93, begelman_98, giannios_06}. Magnetic reconnection is an inherently time-dependent, highly dynamic process as solar observations and recent numerical simulations have revealed \citep[e.g.][]{lin_05, kliem_10}. For the highly conducting plasma of blazar jets, the reconnection current sheets are susceptible to tearing instabilities that lead to their fragmentation in a chain of plasmoids (or, magnetic islands), i.e., regions containing magnetic fields and energetic particles \citep[e.g.][]{loureiro_07, daughton_07, bhattacharjee_09, loureiro_12}. The plasmoids grow rapidly through mergers before leaving the reconnection region. 
Occasionally, plasmoids can undergo significant growth to a sizable fraction of the reconnection region, forming ``monster'' plasmoids \citep{uzdensky_10}. The time-dependent aspects of reconnection may prove to be crucial in understanding blazar flares (in terms of energetics and timescales) as shown by \citet{giannios_09} 
and \citet{giannios_13}.

{The most fundamental way to capture the formation, dynamics, particle and magnetic energy content of plasmoids in reconnection layers is by means of fully kinetic particle-in-cell (PIC) simulations. PIC simulations of reconnection} have been recently extended to the relativistic regime 
of $\sigma \gtrsim 1$ that is relevant to blazars \citep{zenitani_01, guo_14, ss_14, nalewajko_15, sironi_15, kagan_16, werner_16}. In particular, \citet{sironi_15} -- henceforth SPG15, showed using two-dimensional (2D) PIC simulations\footnote{\citet{ss_14} showed that the long term evolution of {relativistic reconnection}, including particle acceleration, proceeds similarly in 2D and 3D.} of electron-positron and electron-ion plasma that relativistic reconnection can satisfy all the basic conditions for the blazar emission: efficient dissipation, extended particle distributions, and rough equipartition between particles and magnetic field in the emitting region, thus supporting  the view that the dissipated energy appears in bursts associated with individual plasmoids in the layer \citep{giannios_13}. The statistical properties of the plasmoid chain, such as size distribution, particle and magnetic energy content,  were recently presented in \citet{sironi_16} (hereafter, SGP16). By employing 2D PIC 
simulations in pair plasmas extended to unprecedentedly long time and length 
scales, SGP16 were able to assess the {basic} properties of the reconnection layer, {namely the particle distributions, the geometry, and the motion of individual plasmoids}, as a function of the system size. This allows us to extrapolate these results from the small plasma scales of PIC simulations to the macroscopic scales relevant for the blazar emission.

Aim of the present study is to  incorporate the physics that describes the plasmoid, e.g.  growth rate of a plasmoid,  magnetic field strength, and injection rate of particles, into a model for its emission. To achieve our goal, we combine (i) recent results from PIC simulations as presented in SGP16 with (ii) the kinetic equation for the evolution of {the distribution of radiating particles} and their synchrotron, synchrotron self-Compton (SSC) emission. Our approach provides physical insight on the basic properties of the plasmoid-powered flares, such as their rise timescale, {and leads} to several robust predictions. We show that synchrotron and SSC flares of duration of several hours--days are powered by the largest and {slowest} plasmoids that form in a reconnection layer. Smaller and {faster} plasmoids, on the other hand, produce flares of similar, {or even higher,} peak luminosity but with sub-hour duration. 

This paper is structured as follows. In \S\ref{sec:reco} we summarize the basic results of the PIC simulations presented in SGP16. In \S\ref{sec:model} we present an analytical model for the electron distribution in the blob and its motion. {In \S\ref{sec:properties} we present the basic properties of flares produced by individual plasmoids and we apply our model to the blazar emission in \S\ref{sec:results}. 
In \S\ref{sec:observables} we present analytical expressions for several observables of flares and we continue with some indicative examples of plasmoid-powered flares in \S\ref{sec:examples}. We discuss various aspects of our model in \S\ref{sec:discussion} and we conclude in \S\ref{sec:summary} with a summary of our results.} 
\section{Basic results from PIC simulations}
\label{sec:reco}
\subsection{Summary of SGP16 results}
In SGP16 we employed a suite of large-scale 2D PIC simulations in electron-positron plasmas to demonstrate that relativistic magnetic reconnection can naturally account for the formation of quasi-spherical plasmoids filled with high-energy particles and magnetic fields. The simulations extended to unprecedentedly
long temporal and spatial scales, thus allowing us to capture the asymptotic physics of plasmoid formation independently of
the initial setup.  We showed that the plasmoids are continuously generated as a self-consistent by-product of the reconnection process and their most important properties are: 
\begin{enumerate}
 \item there is rough equipartition between particle kinetic energy density and magnetic field energy density;
 \item the comoving particle density and magnetic field strength of {each} plasmoid remain approximately constant during its growth;
 \item the plasmoids grow in size at $\sim 0.1$ of the speed of light (i.e., at about half of the reconnection inflow rate), with most of the growth happening while they are still non-relativistic;
 \item their growth is suppressed once they get accelerated to relativistic speeds, up to a terminal four-velocity $\sqrt{\sigma}\, c$;
 \item the width $w^{\dpr}$ of the largest (monster) plasmoids is $w^{\dpr}\sim 0.2\ell^\prime$, independently of the size of the reconnection layer $2 \ell^\prime$; 
 \item {plasmoids with sizes much larger than the characteristic plasma scales\footnote{The characteristic plasma scale is the plasma skin depth defined as $c/\omega_{\rm p}$, where $\omega_{\rm p}=eB_0^{\prime}/m c \sqrt{\sigma}$ is the plasma frequency. For the astrophysical application of this paper $B_0^\prime\sim$ 1~G and  $\sigma\sim 10$, which results in skindepths of $\sim 10^6 {\rm cm}\ll \ell^\prime \sim 10^{16}$~cm; see next sections.} contain isotropic particle distributions (see also Table~ 1, Figs.~7 and 12 in SGP16)}; and  
 \item the typical recurrence interval  of the largest plasmoids is $\sim 2.5\ell^\prime/c$, while smaller plasmoids are more frequent.
 \end{enumerate}

\section{Flares from plasmoids: an analytical model}
\label{sec:model}
In this section we incorporate the results from PIC simulations regarding the dynamical evolution of the plasmoids into an analytical  model for the plasmoid emission. We first present the basic assumptions that enter our calculations (\S\ref{sec:assumptions}) and continue with a description of the plasmoid and particle evolution (\S\ref{sec:motion}-\ref{sec:elec-evol}). The characteristic properties of plasmoid-powered flares are presented in \S\ref{sec:properties}.

\subsection{Assumptions}
\label{sec:assumptions}
\begin{enumerate}
 \item The plasmoid is homogeneous with constant particle density and  magnetic field strength throughout its volume; PIC simulations show that plasmoids have a structure, namely the particle kinetic and magnetic energy densities peak at the core of the plasmoid and decrease towards its outer parts (see e.g. Fig.~A1 in SGP16). Although an inhomogeneous emission model is more realistic, this will not significantly alter the main conclusions of this work regarding the multi-wavelength spectra and light curves, since we use the volume-averaged properties of the plasmoid as determined by the PIC simulations.  
 However, the inhomogeneous structure of the plasmoids is crucial for calculating the polarization signatures and we plan to investigate this in the future. 
 \item The plasmoid in its rest frame is a sphere; although this is a good approximation for plasmoids studied in 2D simulations (see e.g. Fig. 5 in SGP16), the plasmoids seen in 3D simulations of relativistic reconnection are best described as ellipsoids  elongated along the direction of the electric current. Since 3D {PIC} studies of the plasmoid chain formation are still premature, we will adopt the results of 2D simulations. 
 \item The particle distribution contained in a plasmoid is isotropic for plasmoids of all sizes; PIC simulations show that anisotropy is present in the smallest plasmoids with sizes a few tens of the plasma scale, whereas particles confined in the largest plasmoids have  approximately isotropic  distributions. Given that  macroscopic plasmoids responsible for the blazar flares are much larger than the plasma scales of the problem \citep{giannios_13}, it is safe to assume that they are  characterized by quasi-isotropic particle distributions. Regardless, even if the anisotropy is present in the early phases of growth, when the plasmoid is still small, we show that the emission produced at this  stage is a negligible fraction of the emission at the peak time of a blazar flare. Thus, our assumption would not introduce any substantial errors in our estimates of peak luminosity and flare 
timescales. 
 \item The particles are accelerated to an extended power-law distribution; the minimum and maximum Lorentz factors of the electron distribution are estimated for an electron-proton plasma. Although PIC simulations in SGP16 were performed for an electron-positron plasma, we argue that all their basic results will hold for electron-proton reconnection as well, since for $\sigma\gg1$ the field dissipation results in nearly equal amounts of energy transferred to protons and electrons (see SPG15). So, the mean energy per particle of the two species is nearly the same, as it is the case for an electron-positron plasma {(for more detailed discussion, see Sect.~3.3).}
 \end{enumerate}
 
\begin{figure}
\centering
\includegraphics[height=0.48\textwidth]{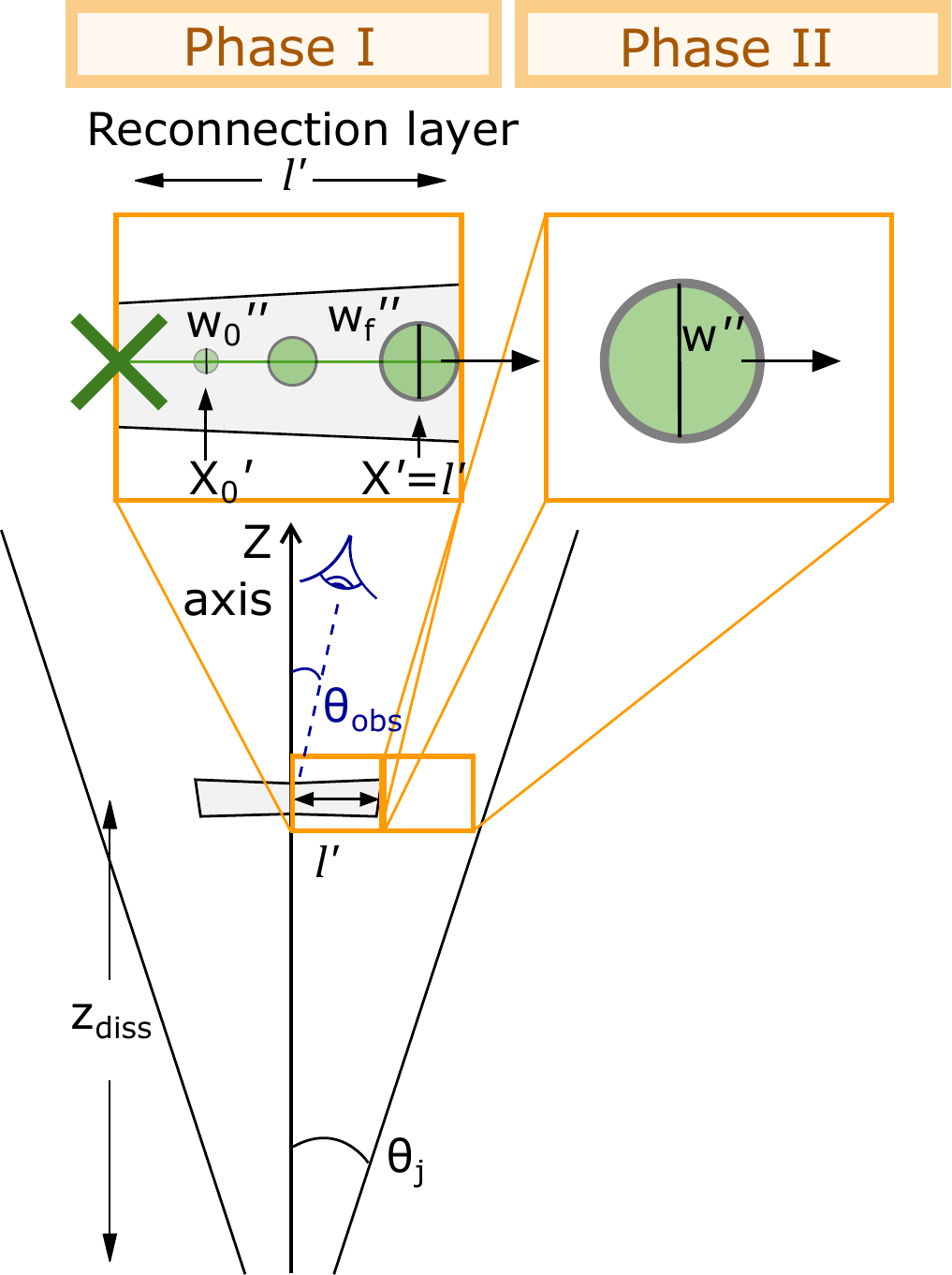}
 \caption{Sketch of a reconnection layer formed at a distance $z_{\rm diss}$ in the jet. The angle between the observer's line of sight and the jet axis is 
 $\thobs$ while $\theta_{\rm j}$ is the jet's opening angle. A plasmoid formed close to the central X-point of the current  sheet grows in size as it moves along the current sheet ({\sl Phase I}) and accumulates particles. At the same time it may accelerate
 from non-relativistic to relativistic speeds, reaching a terminal velocity that is close to the Alfv\'{e}n speed. The injection of particles ceases when the plasmoid leaves the current sheet ({\sl Phase II}). The growth of the plasmoid size in this phase is caused by expansion in, e.g., the under-pressured  surrounding jet plasma.}
 \label{fig:sketch}
\end{figure}
 \begin{figure*}
 \centering 
 \includegraphics[width=0.33\textwidth]{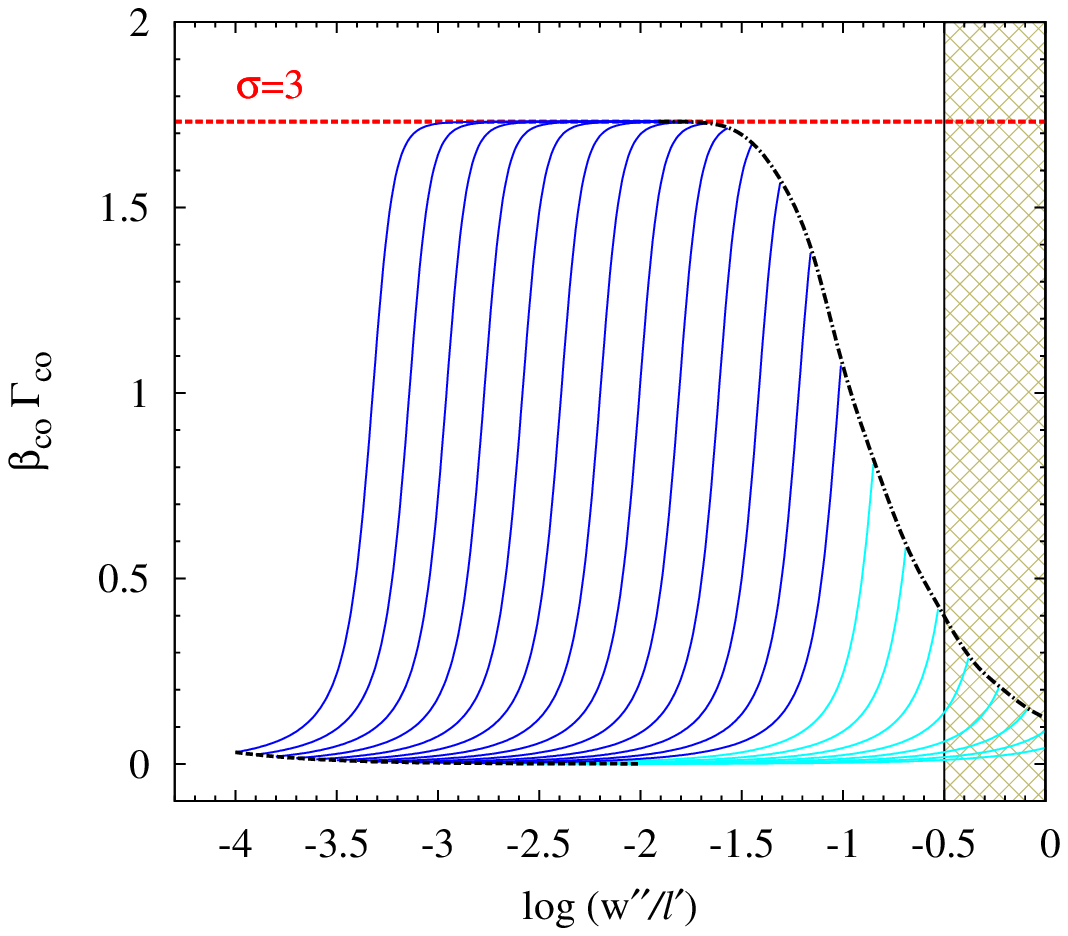}
 \includegraphics[width=0.33\textwidth]{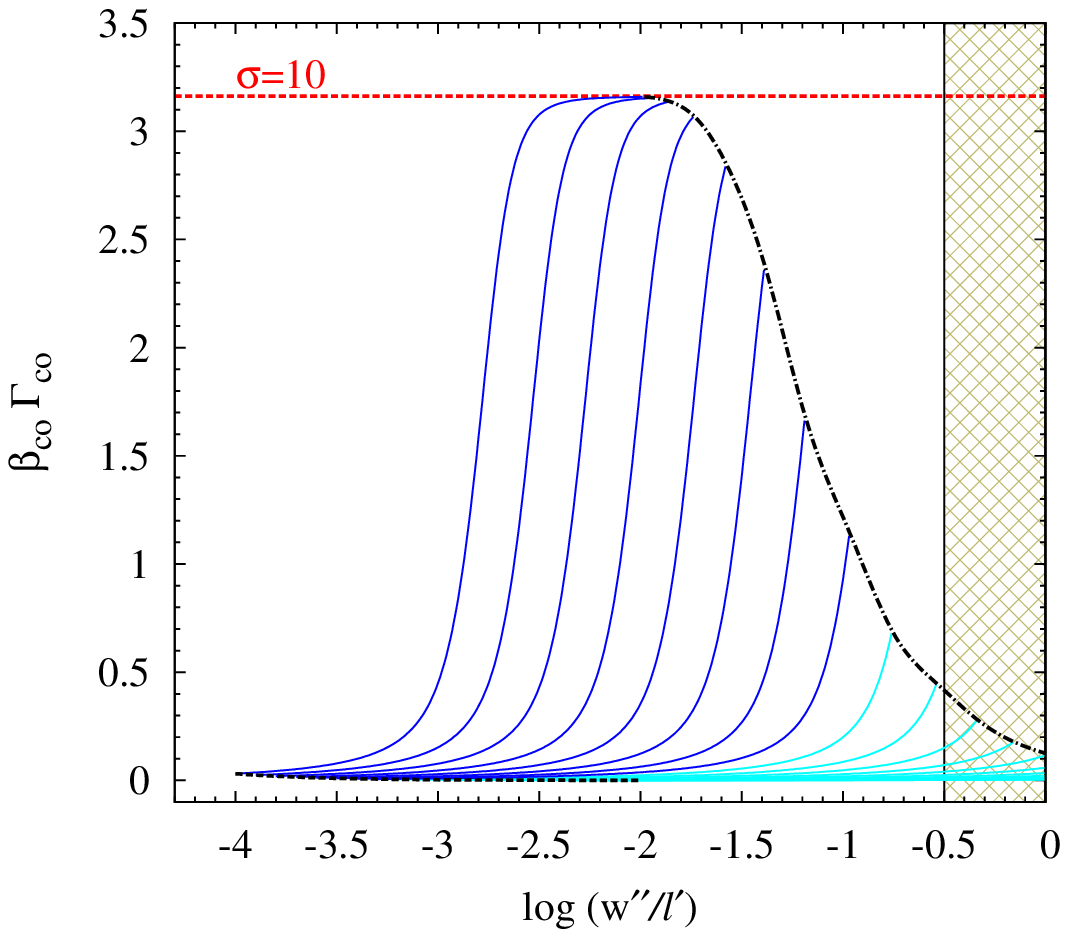}
 \includegraphics[width=0.33\textwidth]{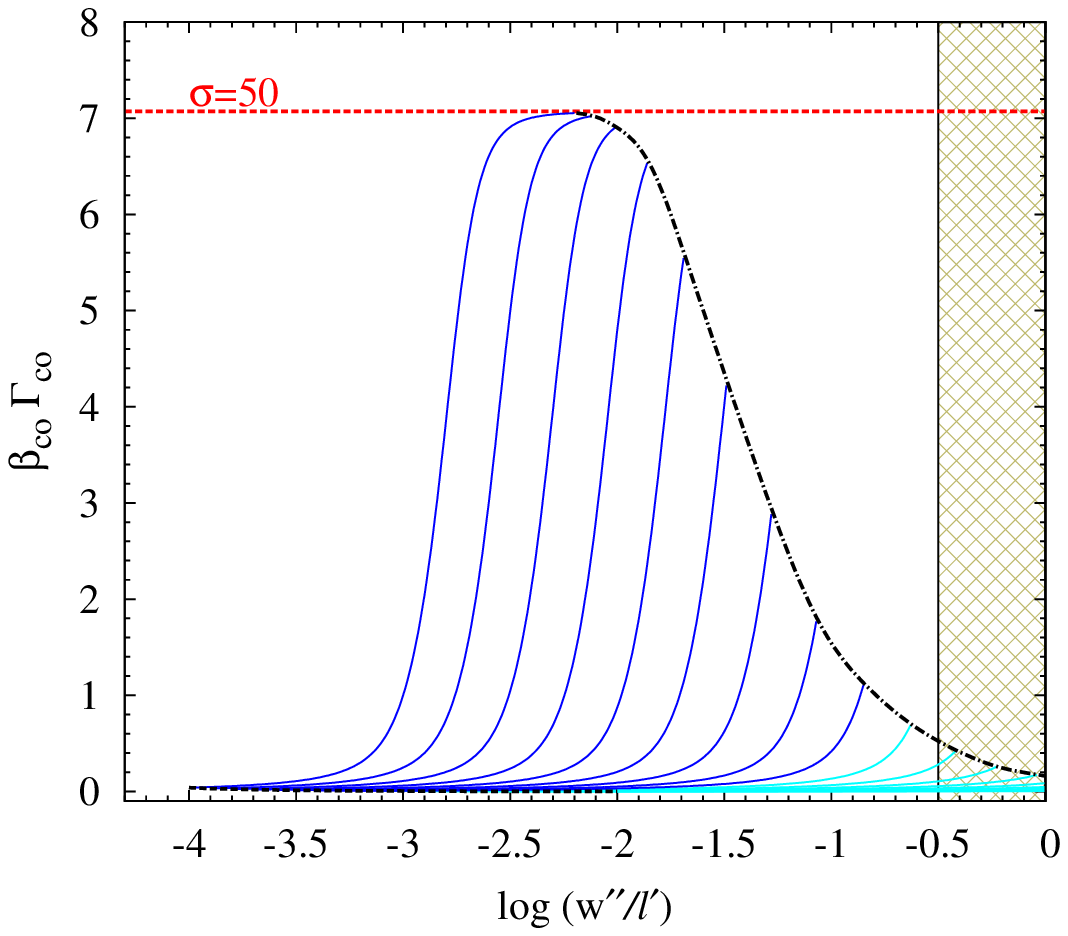}
 \caption{Plasmoid momentum $\bco \Gco$ as a function of {its width normalized to the half-length of the layer,} $w^{\dpr}/\ell^\prime$, for different magnetizations marked on the plot. Each coloured line denotes the evolutionary path of a single plasmoid forming at  $X^\prime_0=0.01\ell^\prime$ with initial width $w^{\dpr}_0$ (black dashed line, nearly horizontal at $\Gco \bco \simeq 0$) and exiting the reconnection layer ($X^\prime=\ell^\prime$) with $w^{\dpr}_{\rm f}$ (black dashed-dotted line). The hatched region corresponds to plasmoids with $w^{\dpr}_{\rm f} > 0.3\ell^\prime$ that are very rare. The paths of plasmoids that leave the layer with $\bco \Gco >1$ are shown with blue coloured lines. Cyan coloured lines correspond to plasmoids that exit the layer while being non-relativistic. The red coloured dashed line marks the asymptotic value ($\sqrt{\sigma}$) of the plasmoid dimensionless four-velocity.  Other parameters used are: $\vg=0.06$, $\vacc=0.12$ (left panel),  $\vg=0.08$, $\vacc=0.12$ (middle 
panel) and $\vg=0.1$, $\vacc=0.15$ (right panel),  as informed by the SGP16 PIC simulations. }
 \label{fig:motion1}
\end{figure*}
\subsection{Plasmoid motion and Doppler boosting}
\label{sec:motion}
We assume that at a distance $z_{\rm diss}$ from the base of the jet (see Fig.~\ref{fig:sketch}) a current sheet of length $2 \ell^\prime$ is formed  (see \S\ref{sec:results} for the large-scale jet model and the motivation for $\ell^\prime, z_{\rm diss}$ parameters). This is  embedded in a relativistic flow with bulk Lorentz factor $\Gj$.  Let us consider  a plasmoid that forms close to the central X-point of the current sheet, i.e. at $X^\prime_0 \ll \ell^\prime$,  with initial width $w^{\dpr}_0$.\footnote{There are three reference frames that are of relevance in our study: (i) the rest frame of the plasmoid (double-primed quantities), (ii) the rest frame of the jet (primed quantities), and (iii) the observer's frame (unprimed quantities).} 
This moves along the current sheet with a speed $\bco$ (in units of the speed of light) as measured in the jet's rest frame and Lorentz factor $\Gco \equiv \left(1- \beta_{\rm co}^2\right)^{-1/2}$. Based on the results presented in SGP16 (in particular, see Fig. 10 and eq.~(11) therein), the plasmoid's momentum as measured in the jet frame is related to $X^\prime/w^{\dpr}$ as 
\eqb
\bco \Gco \approx f\left(\frac{X^\prime}{w^{\dpr}}\right)\equiv \sqrt{\sigma} \tanh\left(\frac{\vacc}{\sqrt{\sigma}}\frac{X^\prime-X^\prime_0}{w^{\dpr}}\right), \, {\rm for} \, w^{\dpr} \ge w^{\dpr}_0,
\label{eq:PIC}
\eqe
where $X^\prime$ is the position along the current sheet and $\vacc$ is a dimensionless number determined numerically that quantifies 
the acceleration rate of the plasmoid and is approximately independent of the magnetization $\sigma$ (i.e., $\vacc=0.12-0.15$ for $\sigma=3-50$). Exactly at $X^\prime_0$ the plasmoids are formed with initial momentum $\bco \Gco \ll 1$ that is not included in eq.~(\ref{eq:PIC}) for simplicity. 

Based on eq.~(\ref{eq:PIC}), two asymptotic regimes of the plasmoid's motion can be identified:
\begin{itemize}
 \item $\bco \Gco \approx \vacc (X^\prime/w^{\dpr})$, \, for $X^\prime/w^{\dpr} \ll 30 (\sqrt{\sigma_1}/\beta_{\rm acc,-1})$\\ 
 \item $\bco \Gco \approx  3 \sqrt{\sigma_1}$, \, for  $X^\prime/w^{\dpr} \gg 30 (\sqrt{\sigma_1}/\beta_{\rm acc,-1})$,
\end{itemize}
where  $\sigma=10\, \sigma_1$ and $\beta_{\rm acc}  = 10^{-1}\,\beta_{\rm acc,-1}$. 

As the plasmoid moves along the current layer it grows in size, mainly through mergers, with a rate $\vg$ that is a significant fraction of the speed of light. SGP16 determined that $\vg \sim 0.06, 0.08$, and 0.1 for $\sigma=3, 10$, and 50 respectively. These values of the growth rate are appropriate for plasmoids whose speed is not too close to the  Alfv{\'e}n speed $\beta_{\rm A}c=\sqrt{\sigma/(1+\sigma)}c$. To account {for the fact that the plasmoid growth gets slower as} $\bco \rightarrow \beta_{\rm A}$  (see Fig. 8 in SGP16) we replace $\vg$ by a suppressed growth rate. This is modelled by $\vg [1+2\tanh(2\bco/\beta_{\rm A})]^{-1}$, so that the asymptotic growth rate is $\vg/3$.\footnote{We have checked that a different form of the suppression factor does not change our main conclusions.}  Thus, the equation that governs the plasmoid growth in its rest frame is
\eqb
\label{eq:dw}
{\rm d}w^{\dpr} = \frac{\vg c  {\rm d}t^{\dpr}}{g\left(\frac{X^\prime}{w^{\dpr}}\right)} = \frac{\vg c {\rm d}t'}{\Gco\left(\frac{X^\prime}{w^{\dpr}}\right) g\left(\frac{X^\prime}{w^{\dpr}}\right)},
\eqe
where  $g(x)\equiv 1+2\tanh\left(2\bco(x)/\beta_{\rm A}\right)$.  In the numerical calculations that follow the modified $\vg$ will be used, whereas $\vg$ will be kept constant in the analytical calculations. This introduces only a small error while greatly simplifies the analytical calculations.  Solving for $\bco={\rm d}X^\prime/c\,{\rm d}t^\prime$ from eq.~(\ref{eq:PIC}) we find
\eqb
\label{eq:dX}
{\rm d}X^\prime = c \bco {\rm d}t^\prime = c {\rm d}t^\prime \frac{f\left(\frac{X^\prime}{w^{\dpr}}\right)}{\sqrt{1+f^2\left(\frac{X^\prime}{w^{\dpr}}\right)}}.
\eqe
The equation that relates the size of the plasmoid with its position $X^\prime$ along the current sheet is found by combining eqs.~(\ref{eq:dw}) and (\ref{eq:dX})
\eqb
\label{eq:dXdw}
\frac{{\rm d}X^\prime}{{\rm d}w^{\dpr}} = \vg^{-1}  g\left(\frac{X^\prime}{w^{\dpr}}\right)f\left(\frac{X^\prime}{w^{\dpr}}\right).
\eqe
The rate of momentum change at the initial stages of a plasmoid's evolution can be written as  
${\rm d}(\bco \Gco)/{\rm d}t^{\dpr}= c(\vacc-\vg)\Gco\bco/w^{\dpr}$, where eqs.~(\ref{eq:PIC}) and (\ref{eq:dXdw}) have been used and $g\approx 1$. {Thus, the plasmoid will accelerate, for $\vacc > \vg$ as found in PIC simulations.}  

The momentum $\bco \Gco$ as a function of $w^{\dpr}/\ell^\prime$ is plotted in Fig.~\ref{fig:motion1} for fiducial plasmoids and different magnetizations marked on the plots.  Each coloured line denotes the evolutionary path of a single plasmoid forming at $X^\prime_0=0.01\ell^\prime$ with initial width $w^{\dpr}_0$ (black dashed line) and exiting the reconnection layer with $w^{\dpr}_{\rm f}$ (black dashed-dotted line). The hatched region corresponds to extremely rare, big plasmoids with $w^{\dpr} > 0.3 \ell^\prime$. Large plasmoids ($\wf\gtrsim 0.1 \ell^\prime$) leave the reconnection layer with a mildly relativistic speed $\bco \Gco\lesssim 1$, whereas smaller plasmoids ($\wf \lesssim 0.04 \ell^\prime$) exit with the asymptotic (dimensionless) four-velocity $\sqrt{\sigma}$ (see also \citet{lyubarsky_05}). As case studies of a large/slow plasmoid and a small/fast plasmoid we, respectively, adopt $\wf =0.2\ell^\prime$ and $ \wf=0.04\ell^\prime$.

We further assume that the current sheet and, in turn, the plasmoid's direction of motion form an angle $\theta'$ with respect to the jet axis, which for convenience we define as the $z$-axis. Without loss of generality, we adopt the $y-z$ plane to be the plasmoid's plane of motion. The plasmoid's velocity in the observer's frame is defined as $\beta_{\rm p}$ and its components parallel and normal to the jet axis 
are given by
\eqb
\beta_{\rm p, z} = \frac{\beta_{\rm j}+ \beta_{\rm co}\cos\theta'}{1 + \beta_{\rm j}\beta_{\rm co}\cos\theta'}, \quad \beta_{\rm p, y} =\frac{\beta_{\rm co}\sin\theta'}{\Gj(1 + \beta_{\rm j}\beta_{\rm co}\cos\theta')}
\eqe
The corresponding Lorentz factor $\Gp = (1-\beta_{\rm p}^2)^{-1/2}$ is given by
\eqb
\Gp = \Gj \Gamma_{\rm co} \left(1+ \beta_{\rm j}\beta_{\rm co}\cos \theta' \right),
\label{eq:G-plasmoid}
\eqe
while the angle of the plasmoid velocity with respect to the $z$ axis  as measured in the lab frame is 
\eqb
\tan \theta =\frac{\bco \sin\theta'}{\Gj\left(\beta_{\rm j}+\bco\cos \theta' \right)}.
\label{eq:theta}
\eqe
Let the observer's position also lie on the $y-z$ plane and $\thobs$  be the angle between the observer's line of sight (see Fig.~\ref{fig:sketch}) and jet axis and $\omega=\theta-\thobs$ be the angle between the plasmoid's direction of motion and the line of sight. The Doppler factor of the plasmoid is then defined as 
\eqb
\dpl = \frac{1}{\Gp \left(1-\beta_{\rm p} \cos \omega \right)},
\label{eq:dop-plasmoid}
\eqe
where $\cos \omega =\cos \theta\cos\thobs+ \sin \theta\sin\thobs$. The Doppler factor corresponding to the bulk motion of the jet alone is defined as 
\eqb
\delta_{\rm j} = \frac{1}{\Gj\left(1-\beta_{\rm j}\cos\thobs\right)}.
\label{eq:dop-jet}
\eqe
In the extreme case of {\it perfect alignment} among the direction of the plasmoid's motion, the jet's bulk motion and the line of sight, namely $\theta'=\thobs=0$, one derives the maximum Doppler boosting $\dpl \simeq 4 \Gco \Gj$ \citep[see also][]{giannios_09} for relativistic plasmoids. For non-relativistic plasmoids and perfect alignment the Doppler factor simplifies to $\dpl \simeq  \delta_{\rm j} \simeq 2\Gj$. 

\begin{figure}
\centering
\includegraphics[width=0.49\textwidth]{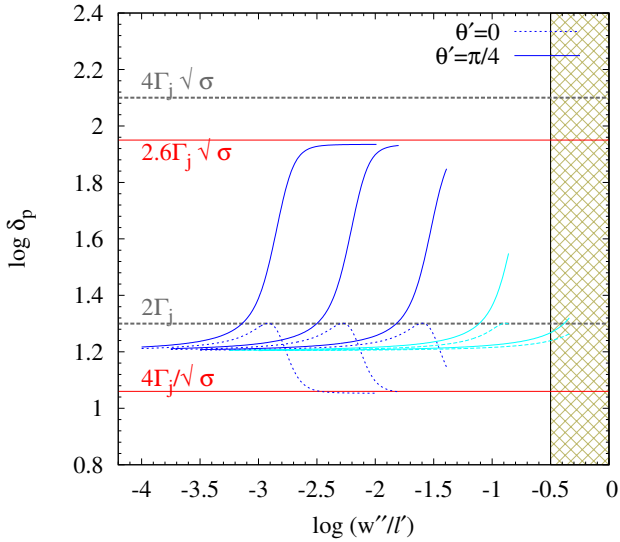}
\caption{Log-log plot of the plasmoid Doppler factor $\dpl$ as a function of its width $w^{\dpr}/\ell^\prime$ for $\sigma=10$, and two choices of the inclination of the layer: $\theta^\prime=0$ (dashed lines) and $\theta^\prime=\pi/4$ (solid lines). Here, the viewing angle is $\thobs=0.5/\Gj$. Blue and cyan coloured lines have the same meaning as in Fig.~\ref{fig:motion1}. The horizontal grey dashed lines indicate the limiting values for $\dpl$ in the case of {\it perfect alignment} (i.e., $\theta'=\thobs=0$) assuming a non-relativistic plasmoid (bottom dashed line) or a relativistic plasmoid moving with the maximum speed (top dashed line). The solid red lines mark the limiting Doppler factor of a relativistic plasmoid for a {\it non-favorable} orientation (i.e., $\thobs=0.5\Gj$, $\theta^\prime=0$;  bottom line) and a {\it favorable} orientation (i.e.,  $\thobs=0.5/\Gj$, $\theta^\prime=\pi/4$; top line). }
\label{fig:dp-thprime}
\end{figure}

The evolution of $\dpl$ as a plasmoid grows in size and accelerates is illustrated in Fig.~\ref{fig:dp-thprime} for $\sigma=10$, two values of $\theta^\prime$ marked on the plot and a viewing angle $\thobs=0.5/\Gj$. Blue and cyan coloured lines have the same meaning as in Fig.~\ref{fig:motion1}. The horizontal grey dashed lines indicate the limiting values for $\dpl$ in the case of perfect alignment (i.e., $\theta'=\thobs=0$) assuming a non-relativistic (bottom dashed line) or a relativistic (top dashed line) plasmoid moving with the Alfv{\'e}n speed ($\bco\Gco\simeq \Gco \simeq \sqrt{\sigma}$). {Fig.~\ref{fig:dp-thprime} shows that the Doppler factor of a plasmoid varies significantly during its growth phase and that it  depends strongly on the combination of $\thobs$ and $\theta^\prime$. The curves $\dpl(w^{\dpr}/\ell^\prime)$ for the two choices of the layer's orientation considered here are not only quantitatively but also qualitatively different. A few remarks on the plot follow.
\begin{itemize}
 \item For a {\it non-favorable} orientation, i.e. $\thobs=0.5/\Gj$ and $\theta^\prime=0$, the plasmoid Doppler factor is written as 
 \eqb
 \label{eq:Doppler-theta0}
 \dpl=2\Gj\Gco(1+\beta_{\rm j}\bco)\left[1+\left(\frac{\Gco(1+\beta_{\rm j}\bco)}{2}\right)^2\right]^{-1}.
 \eqe
There are two regimes of interest in the evolution of a single plasmoid, namely
\begin{enumerate}
 \item the non-relativistic regime, where $\bco\Gco \ll 1$. Equation (\ref{eq:Doppler-theta0}) simplifies then 
 into $\dpl \approx (8/5)\Gj \simeq  16 \Gamma_{\rm j, 1}$,  in agreement with with the asymptotic behavior at small sizes in Fig.~(\ref{fig:dp-thprime}). In fact, $(8/5)\Gj$ is the initial Doppler factor of all plasmoids, since they form  as non-relativistic structures.
 \item the asymptotic and relativistic regime, where $\bco \Gco \approx \Gco \approx \sqrt{\sigma}$. In this case, the Doppler factor is written as 
 $\dpl \approx 4\Gj/\sqrt{\sigma} \simeq 12 \Gamma_{\rm j, 1}/\sqrt{\sigma_1}$. This value is indicated in Fig.~\ref{fig:dp-thprime} by the lower red solid line.
 \end{enumerate}
\item  For a {\it favorable} orientation, i.e. $\thobs=0.5/\Gj$ and $\theta^\prime=\pi/4$, the Doppler factor in the non-relativistic regime is the same as for $\theta^\prime=0$ (see point (i) above), whereas in the relativistic regime it is given by $\dpl \approx 2\Gp$, since $\omega=\theta-\thobs \simeq 0.01$. Thus, $\dpl \approx 2\Gj \Gco (1+\cos\theta^\prime)/[1+(\Gj\Gco\omega(1+\cos\theta^\prime)^2]$, which results in $\dpl \simeq 3.4 \sqrt{\sigma}\Gj/[1+0.3\Gamma_{\rm j, 1}^2 \sigma_1]$ for plasmoids moving with the asymptotic four-velocity (top red solid line in Fig.~\ref{fig:dp-thprime}).
\end{itemize}
 Plasmoids that leave the layer with final sizes $\lesssim 0.03\ell^\prime$ are accelerated close to their asymptotic  four-velocity (see also Fig.~\ref{fig:motion1}) and their Doppler factors depend sensitively on the orientation of the layer and the observer (compare blue solid and dashed lines). The dependence of $\dpl$ on the plasmoid size is determined by both the acceleration and growth processes, while it affects the relative timescales as measured in the observer's frame (see eq.~(\ref{eq:tobs}) below and \S\ref{sec:timescale}). On the contrary, for a fixed viewing angle the orientation of the layer becomes irrelevant for non-relativistic plasmoids and $\dpl \rightarrow 2\Gj$ (cyan solid and dashed lines). 

Fig.~\ref{fig:map-ratio} shows the ratio $\dpl(w^{\dpr}_{\rm f})/\delta_{\rm j}$ for different orientations of the observer ($\thobs$) and the plasmoid's direction  of motion ($\theta^\prime$) with respect to the jet axis. The results for $\sigma=10$ and two different plasmoid sizes are shown in the left ($\wf=0.04 \ell^\prime$) and right ($\wf=0.2\ell^\prime$)  panels. For a small and fast moving plasmoid (left), the plasmoid's Doppler factor may be up to $\sim 5$ times larger than that of the jet's bulk motion alone. For a bigger, yet slowly moving, plasmoid (right), the ratio $\dpl(w^{\dpr}_{\rm f})/\delta_{\rm j}$  varies at most by a factor of $\sim 1.4$ and it is less angle-dependent.
\begin{figure*}
\centering
\includegraphics[width=0.49\textwidth]{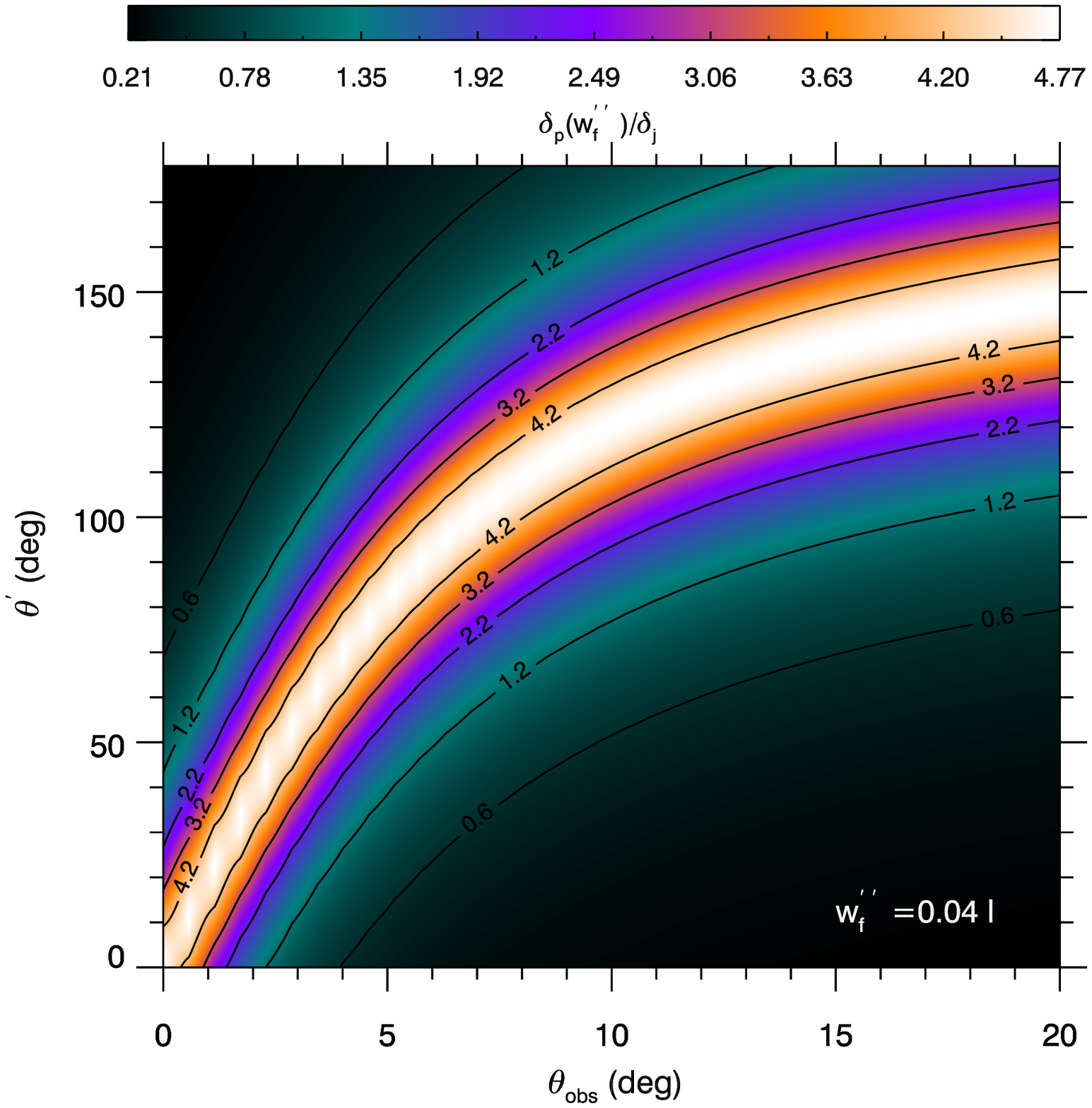}
\includegraphics[width=0.49\textwidth]{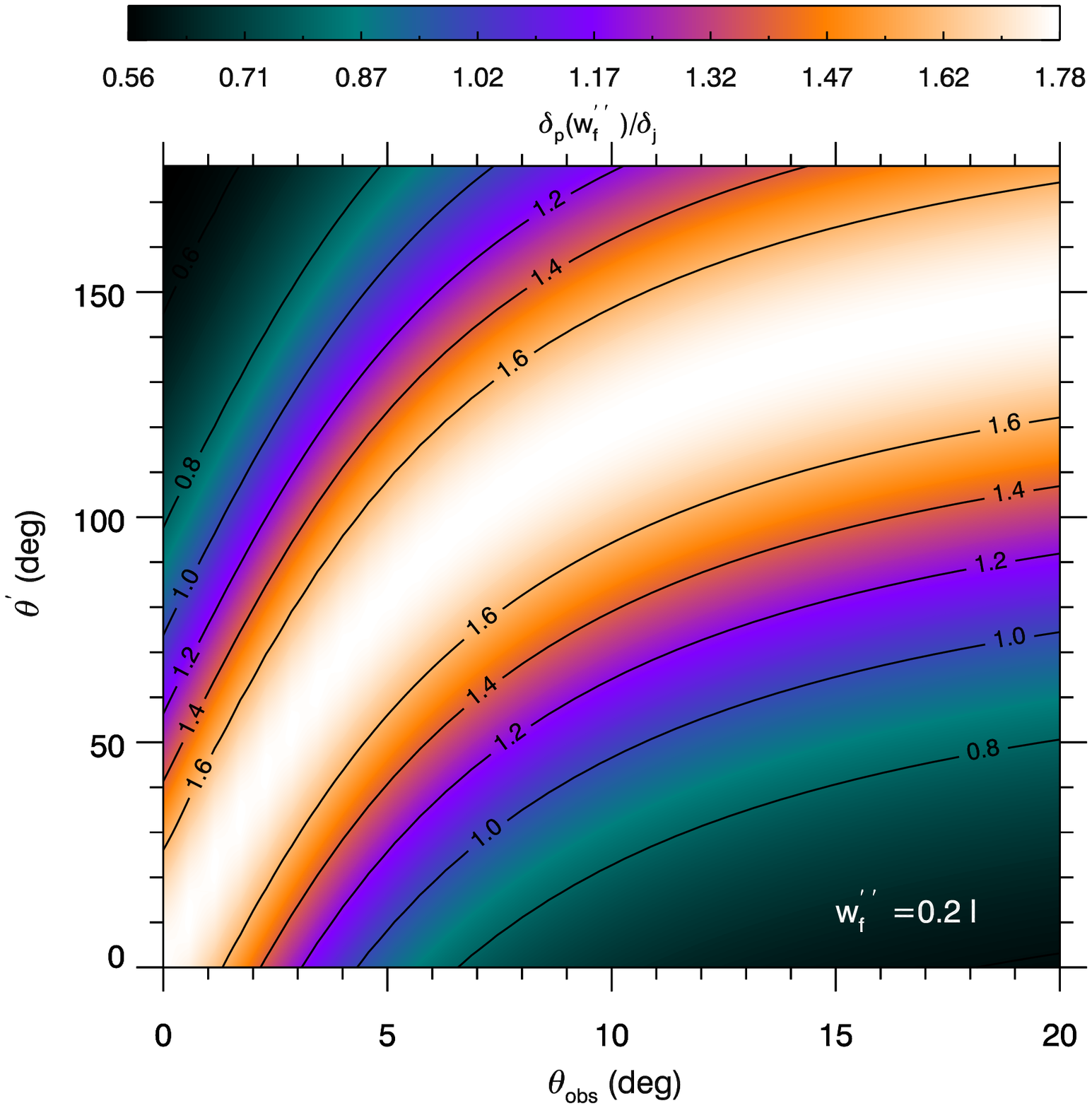}
\caption{Two-dimensional maps of the ratio of $\dpl(w^{\dpr}_{\rm f})/\delta_{\rm j}$ created for $\sigma=10$ and different values of $\theta^\prime$ (i.e., the angle between the jet axis and the plasmoid's direction of motion in the jet frame) and $\thobs$ (i.e., the angle between the jet axis and the observer's line of sight). Overplotted with black lines are contours of fixed ratios  $\dpl(w^{\dpr}_{\rm f})/\delta_{\rm j}$. For illustration purposes, the colour scale in the two plots is different. Left and right panels correspond to plasmoids with $\wf=0.04\ell^\prime$ and $\wf=0.2\ell^\prime$, respectively. The smaller plasmoid is characterized by relativistic motion in the jet frame and its emission is strongly beamed for favorable inclinations of the observer.}
\label{fig:map-ratio}
\end{figure*}

The time for a plasmoid to grow to its size $w^{\dpr}$ as measured in the observer's frame is given by
\eqb 
\label{eq:tobs}
t(1+z)^{-1}= \int \frac{{\rm d}t^{\dpr}}{\dpl\left(w^{\dpr}\right)} = \int^{w^{\dpr}} \!\!\!{\rm d} \tilde{w}\frac{g(\tilde{w})}{c \vg \dpl\left(\tilde{w}\right)},
\eqe
where $X^\prime(w^{\dpr})$ is determined by eq.~(\ref{eq:dXdw}) and $g(x)$ is defined below eq.~(\ref{eq:dw}). For $w^{\dpr}=\wf$, the above expression determines the typical peak time for flares powered by the plasmoid. 

As we detail below (see next section), the calculation of the electron evolution and emission are performed in the rest frame of the plasmoid. If $L^{\dpr}, L^{\dpr}(\nu^{\dpr})$  are, respectively, the bolometric and differential photon luminosities in the plasmoid's frame, their values in the observer's frame are $L=\dpl^4 L^{\dpr}$ and $L(\nu) = \dpl^{3+\alpha}L^{\dpr} (\nu^{\dpr})$, where $\alpha$ is the spectral index. Here, we neglect inverse Compton scattering on external (to the jet) radiation fields which would result in a boosting factor of $\dpl^{4+2\alpha}$ \citep{dermer95}. 

\subsection{Plasmoid evolution}
\label{sec:plasmoid-evol}
The plasmoid evolution can be divided into the following phases (see also Fig.~\ref{fig:sketch}):
\begin{itemize}
 \item {\sl Phase I} that corresponds to the time that the plasmoid spends in the reconnection layer while growing in size, and  
 \item {\sl Phase II} that  begins with the plasmoid leaving the reconnection layer and undergoing expansion in the jet's bulk flow.
\end{itemize}
The modeling of Phase I is entirely based on the results of PIC simulations that we summarize below: (i) the volume of the plasmoid increases as $V^{\dpr}\propto w^{\prime\prime 3}$, (ii)
the magnetic field strength in the rest frame of the plasmoid, $\Bp$, is approximately independent of its size, (iii) the electron number density $n^{\dpr}$ remains approximately constant with respect to the plasmoid's size (see e.g. Fig. 5 in SGP16), (iv) the total number of electrons $N$ increases as $N=n^{\dpr} V^{\dpr} \propto w^{\prime \prime 3}$,
(v) in Phase I the injected particles cool because of radiative energy losses (i.e., synchrotron and inverse Compton scattering); adiabatic cooling is not relevant in this phase, since the number density remains constant, and (vi) the injected particle distribution is a power law from $\gmin$ to $\gmax$ and slope $p$ as determined by PIC simulations. Steep particle spectra with $p>2$ are found for $\sigma \lesssim 10$, whereas $p \lesssim 2$ for  $\sigma \gtrsim 10$ \citep{ss_14,sironi_15,guo_14, werner_16}. 

For electron-proton plasmas with $p>2$ and pair multiplicity $N_{\pm}$, the minimum Lorentz factor is given by
\eqb
 \label{eq:gmin}
 \gmin^{\dpr} \simeq \frac{f_{\rm rec}\sigma}{2N_{\pm}}\frac{p-2}{p-1}\frac{\mpr}{\mel},
 \eqe
where $f_{\rm rec}$ is the fraction of dissipated magnetic energy in relativistic reconnection that is transferred to electrons (see eq.~(3) in SPG15). 
In the derivation above, we assumed that $N_\pm \ll \mpr/\mel$ and we also made use of the fact that the mean energy per particle available for dissipation is $\sigma/2$, where $\sigma/2 \gg 1$ is implicitly assumed.
\citet{sironi_15} showed that $f_{\rm rec}\simeq 15\%$ for $\sigma =3$, but reaches the asymptotic value of 25\% for $\sigma \ge 10$ (see Figs. 3 and 4, therein). These values for $f_{\rm rec}$ were obtained for electron-ion reconnection with  $N_{\pm}\sim 1$. 
If the jet is dominated by pairs in terms of number ($N_{\pm} \gg 1)$, these are expected to pick up most of the dissipated magnetic energy at the expense of the few protons that are present in the jet. Then, the typical value of $f_{\rm rec}$ will be $\sim 0.5$, as appropriate for electron-positron plasmas; nearly half of the available energy will still remain in the magnetic field.
   
 For $\sigma \lesssim 10$ the exact value of $\gmax^{\dpr}$ does not affect the results due to the steepness of the power-law spectrum ($p>2$).  On the contrary, for $\sigma \gg 10$ the energy is carried by the highest energy particles. Thus, the maximum Lorentz factor $\gmax^{\dpr}$ is limited by $\sigma$ \citep{ss_14, guo_15a, werner_16}  and is given by
 \eqb
 \label{eq:gmax}
 \gmax^{\dpr} \simeq \left[ \frac{f_{\rm rec}\sigma}{2 N_{\pm}} \frac{(2-p)\mpr}{(p-1)\mel} \right]^{1/(2-p)},
 \eqe
 where $\gmin^{\dpr}\simeq 1$. These are the gross characteristics of particle distributions in plasmoid-driven reconnection.

\subsection{Electron distribution}
\label{sec:elec-evol}
All the calculations that follow are performed in the rest frame of the plasmoid. 
The evolution of  the electron distribution, $N(\gamma^{\dpr}, t^{\dpr})$, can be described by a partial differential equation, which has the general form
\eqb
\frac{\partial N}{\partial t^{\dpr}} + \frac{\partial}{\partial \gamma^{\dpr}} \left(N\frac{d\gamma^{\dpr}}{{\rm d}t^{\dpr}} \right)
= Q(\gamma^{\dpr}, t^{\dpr}),
\label{general}
\eqe
where $d\gamma^{\dpr}/{\rm d}t^{\dpr} < 0$ is the total energy loss rate and $Q(\gamma^{\dpr}, t^{\dpr})$ is the injection rate of particles with Lorentz factors between $\gamma^{\dpr}, \gamma^{\dpr} + {\rm d}\gamma^{\dpr}$. The term that describes the escape of particles from the plasmoid (i.e., $\propto N/t_{\rm esc}^{\dpr}$) is omitted from eq.~(\ref{general}), since particles with $\gamma^{\dpr} \le \gmax^{\dpr}$ are confined in the blob (see also SGP16, for details). 
\subsubsection*{Phase I}
We express first eq.~(\ref{general}) in terms of the plasmoid width $w^{\dpr}$ using the relation ${\rm d}w^{\dpr}= \vg c {\rm d}t^{\dpr}$\footnote{The growth speed is assumed to be constant in all analytical calculations, i.e, without the suppression factor $g(X^\prime/w^{\dpr})$.}
\eqb
\label{kinetic-1}
\frac{\partial N_{\rm I}}{\partial w^{\dpr}} - \ksone\frac{\partial}{\partial \gamma^{\dpr}} \left(N_{\rm I}  \gamma^{\dpr 2} \right) = Q_{\rm I}(\gamma^{\dpr}, w^{\dpr}),
\eqe
where the subscript `I' is used to remind of Phase I and 
\eqb
\label{ks1}
\ksone = \frac{\sth B_{\rm p}^{ \dpr 2}}{6 \pi \mel c^2 \vg}.
\eqe
Inverse Compton cooling on a fixed photon target field could be easily incorporated in the above expression as long as the  scatterings take place
in the Thomson regime. We also remark that our analytical calculations presented in this section and Appendix~\ref{sec:app0} do not apply to cases where SSC cooling  dominates the electron energy losses \citep[see e.g.][]{schlickeiser09}. The injection rate for $w^{\dpr}<\wf$ may be written as 
\eqb
\label{Q1}
Q_{\rm I} = \frac{\pi}{2} w^{\dpr 2} n^{\dpr} f_{\rm p} \gamma^{\dpr -p} H[\wf-w^{\dpr}] H[\gamma^{\dpr}-\gmin^{\dpr}] H[\gmax^{\dpr}-\gamma^{\dpr}],
\eqe
where  $f_{\rm p}= (p-1)/(\gmin^{\dpr -p+1}-\gmax^{\dpr -p+1})$ and the normalization of $Q_{\rm I}$ is determined by the condition ${\rm d}N_{\rm I}/{\rm d}w^{\dpr} =  (\pi/2) n^{\dpr} w^{\dpr 2}$.
The solution to eq.~(\ref{kinetic-1}), $N_{\rm I}(\gamma^{\dpr}, w^{\dpr})$, is derived in Appendix~\ref{sec:app0}.
\subsubsection*{Phase II}
In contrast to Phase I, this evolutionary stage  is not benchmarked with PIC simulations. In the analytical treatment (see Appendix~\ref{sec:app0}) we assume that the injection of particles  ceases abruptly when the plasmoid leaves
the layer, i.e. $X^\prime=\ell^\prime$. A gradual cessation of particles is used though in the numerical examples presented in \S\ref{sec:examples}. By this time, the plasmoid has grown up to a size $w^{\dpr}_{\rm f}$ that can be determined
from eq.~(\ref{eq:dXdw}) given an initial condition $X^\prime(w^{\dpr}_0)=X^\prime_0$. 
In Phase II particles undergo synchrotron and adiabatic losses. Because of the expansion of the plasmoid the strength of the magnetic field is
expected to decrease, i.e. $B^{\dpr} \propto w^{\dpr -q}$. The exact value of the exponent $q$ depends upon the magnetic field topology in the plasmoid. 
For example, if $B^{\dpr}$ is turbulent it can be treated as a relativistic fluid with adiabatic index $4/3$, leading to $q=2$. In addition, the plasmoid's expansion is expected to  depend on the properties of the jet's bulk flow, such as pressure and density profiles. In order to keep our formalism as general as possible we model the expansion rate and magnetic field strength as power laws of the size $w^{\dpr}$:
\eqb
\label{vel-exp}
\vexp & = & \vexpo \left(\frac{w^{\dpr}}{\wf}\right)^a, \\
B^{\dpr} & = & B^{\dpr}_{\rm p} \left(\frac{\wf}{w^{\dpr}}\right)^{q}, \ q>0.
\eqe
where $\beta_{\rm exp}$ is the expansion velocity and $a$ is a free parameter that could be 0 for a constant expansion rate or 
$a\lessgtr0$ for an accelerating/decelerating expansion of the plasmoid. The kinetic equation that describes the evolution 
of electrons  during {\sl Phase II} is written as 
  \eqb
 \label{kinetic-2}
 \frac{\partial N_{\rm II}}{\partial w^{\dpr}} + \frac{\partial}{\partial \gamma^{\dpr}} \left(N_{\rm II} \frac{{\rm d}\gamma^{\dpr}}{{\rm d}w^{\dpr}}\right) = Q_{\rm II}(\gamma^{\dpr}, w^{\dpr}),
 \eqe
where 
 \eqb
 \label{char2}
 \frac{{\rm d}\gamma^{\dpr}}{dw^{\dpr}} = -\frac{\gamma^{\dpr}}{w^{\dpr}} -\kstwo \frac{\gamma^{\dpr 2}}{w^{\dpr 2q+a}},
 \eqe
 with the coefficient $\kstwo$ defined as 
 \eqb
 \kstwo = \frac{\sth B_{\rm p}^{\dpr 2} w_{\rm f}^{\dpr 2q+a}}{6 \pi \mel c^2 \vexpo}.
 \label{ks2}
 \eqe
The source term in eq.~(\ref{kinetic-2}) may be written as 
\eqb
Q_{\rm II}(\gamma^{\dpr}, w^{\dpr}) = N_{\rm I}(\gamma^{\dpr}, w^{\dpr})\delta \left(w^{\dpr}-\wf \right),
\label{Q2}
\eqe
namely it describes an instantaneous event of particle injection when the source has width $\wf$, carrying the evolution history of the distribution during Phase I. The electron distribution in Phase II is presented in Appendix~\ref{sec:app0}.

\begin{figure*}
\centering
\includegraphics[width=0.33\textwidth]{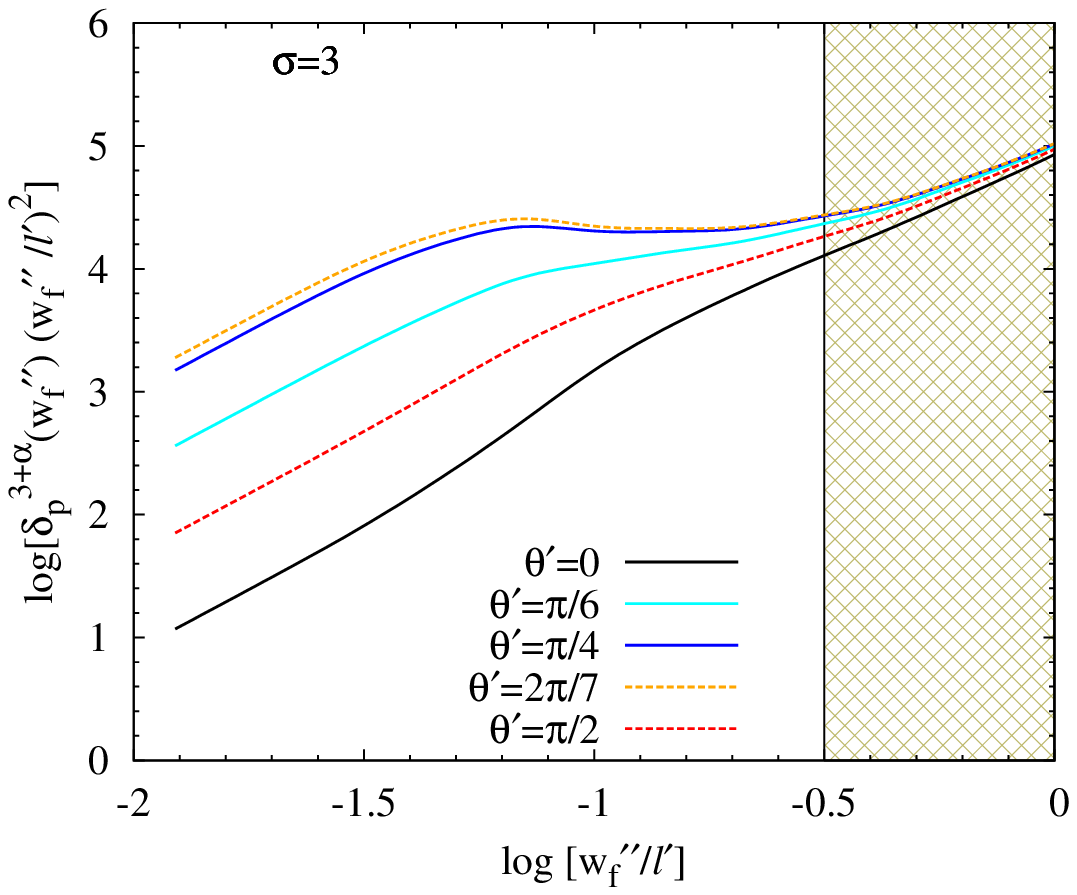}
\includegraphics[width=0.33\textwidth]{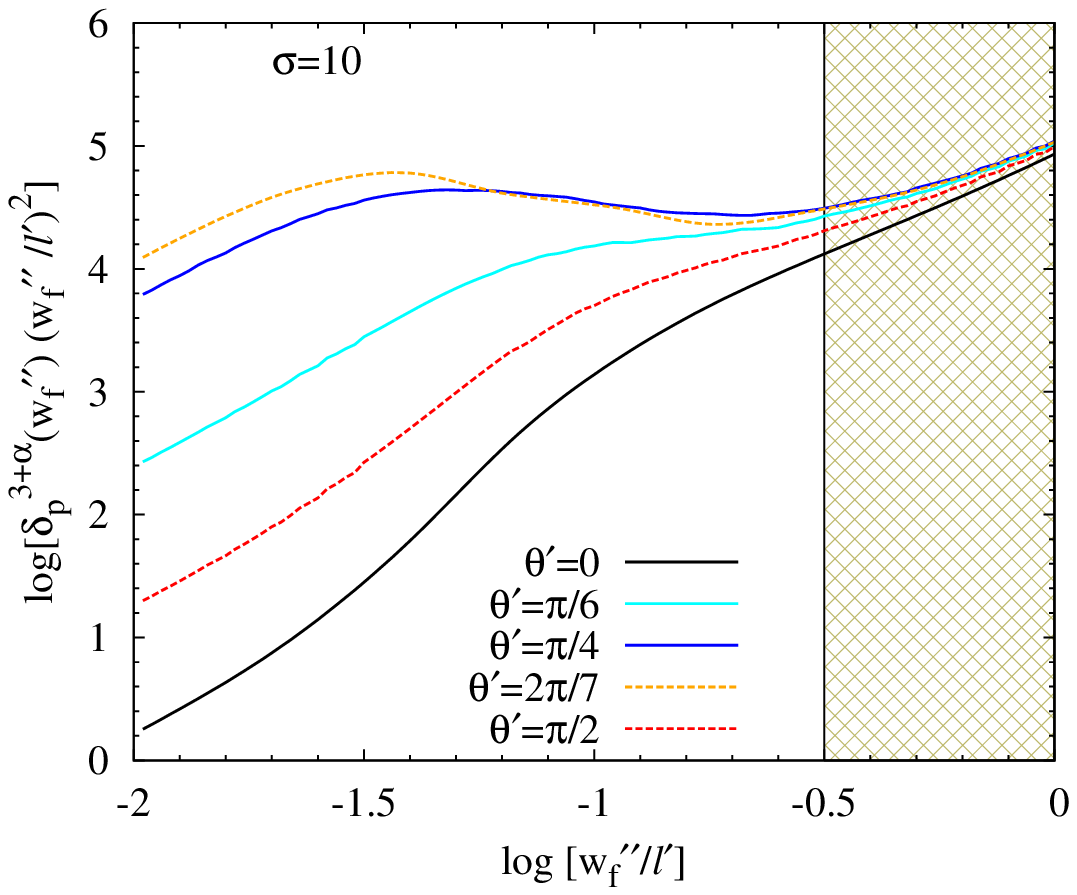}
\includegraphics[width=0.33\textwidth]{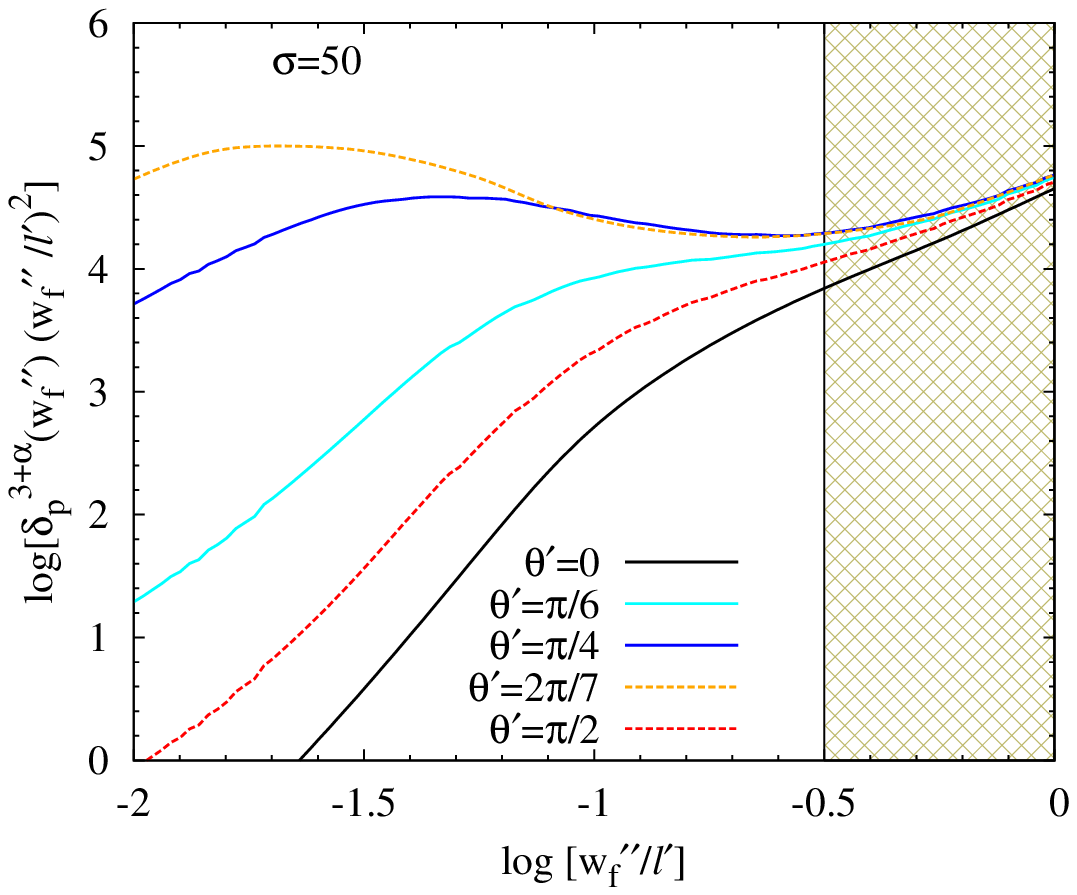}
 \vspace{-0.2in}
 \caption{Log-log plot of the quantity $\left[\dpl(w^{\dpr}_{\rm f})\right]^{3+\alpha}(w^{\dpr}_{\rm f}/\ell^\prime)^2$, which is a measure of the peak luminosity of a plasmoid-powered flare, as a function of $w^{\dpr}_{\rm f}/\ell^\prime$ for $\sigma=3,10, 50$ and various values of $\theta^\prime$ marked on the plot. A fixed viewing angle $\thobs=0.5/\Gj$ has been adopted. The hatched region corresponds to $w^{\dpr}_{\rm f}>0.3 \ell^\prime$. Here, $\alpha=p/2$ is adopted with $p=3, 2$ and 1.5 for $\sigma=3,10$ and 50, respectively  \citep{ss_14}. All other parameters are the same as in Fig.~\ref{fig:motion1}. For favorable orientations of the current sheet small plasmoids power bright flares (see dashed orange and solid blue lines).}
 \label{fig:Lpk_wf}
\end{figure*}
 
As long as SSC cooling does not dominate the electron energy losses \citep[see e.g.][]{schlickeiser09, zacharias_schlickeiser12}, eqs.~(\ref{N1}) and (\ref{N2}) with the accompanying expressions for the cooling break and the lower/upper cutoffs of the distribution can be used directly to calculate the synchrotron and SSC  emission. 

The total number of electrons in Phases I and II can be found by integrating  the expressions (\ref{N1}) and (\ref{N2}), respectively,  over $\gamma^{\dpr}$. As expected, the total number of electrons increases as $w^{\dpr 3}$ during Phase I and  remains constant after that. This, in combination with the constant magnetic field in Phase I and the decaying magnetic field in Phase II, implies that the peak of the emission is expected at the end of Phase I. 
\begin{figure*}
\centering
\includegraphics[width=0.49\textwidth]{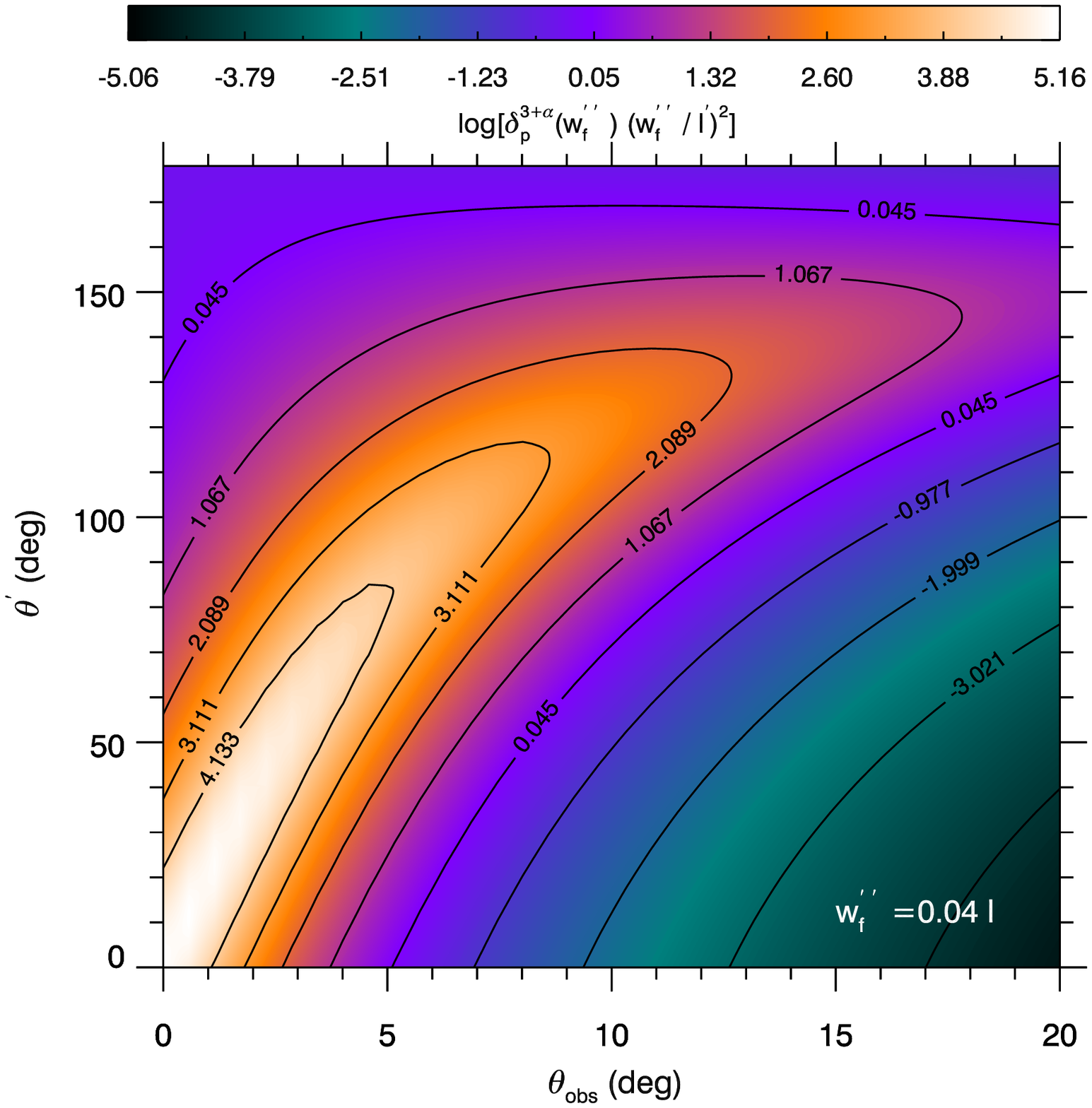}
\includegraphics[width=0.49\textwidth]{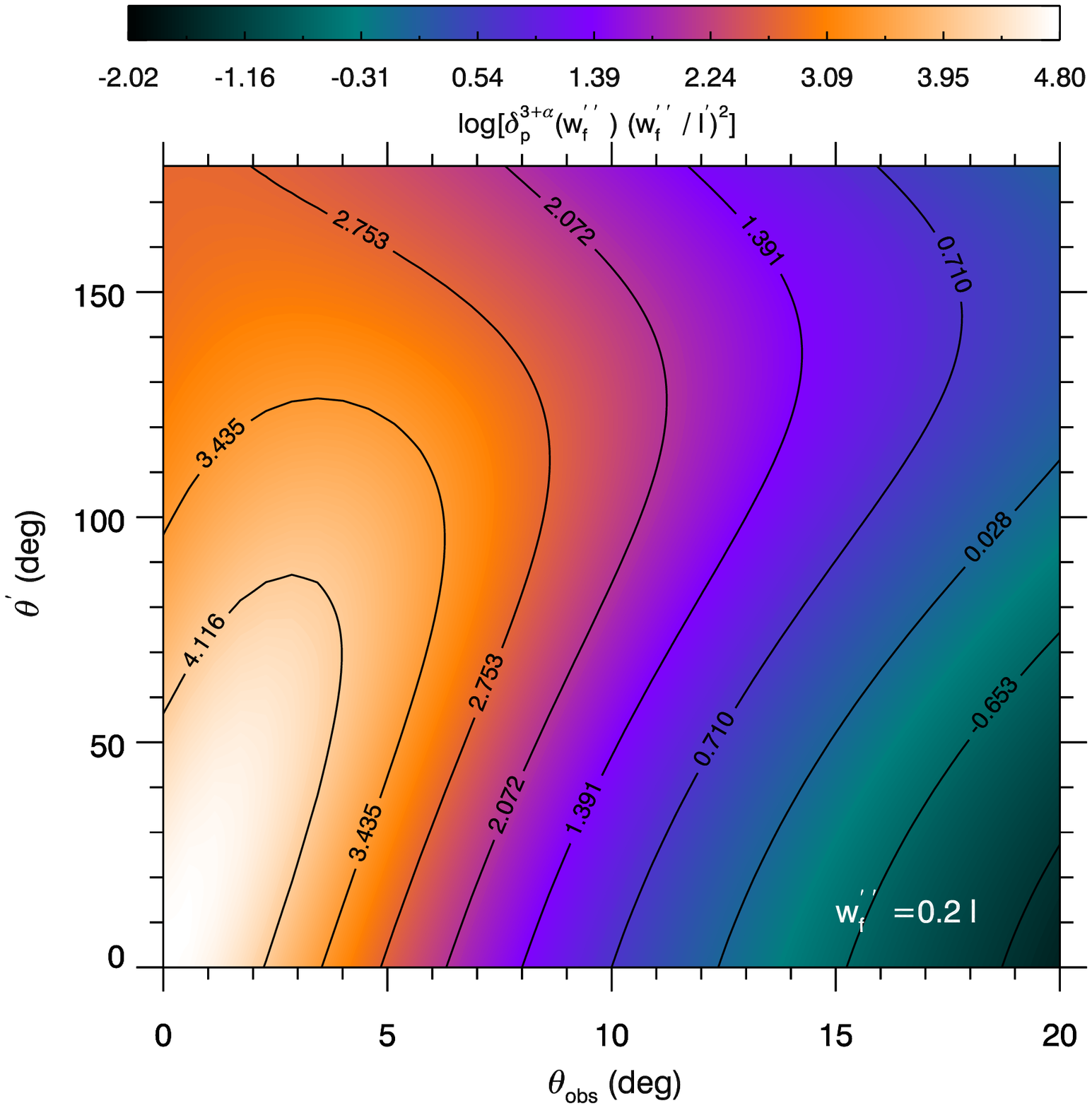}
 \caption{Two-dimensional maps of the quantity $\left[\dpl(w^{\dpr}_{\rm f})\right]^{3+\alpha}(w^{\dpr}_{\rm f}/\ell^\prime)^2$  for a plasmoid with $w^{\dpr}_{\rm f}=0.04\ell^\prime$ (left) and $0.2\ell^\prime$ (right). Here,  $\sigma=10$.  Each map is created for different viewing angles $\thobs$ (in the lab frame) and directions of plasmoid's motion with respect to the jet axis $\theta^\prime$ (in the jet's rest frame).  The colour scale is also different in the two plots. All other parameters are the same as in Fig.~\ref{fig:motion1}.}
\label{fig:map-boost}
\end{figure*} 
\section{Properties of plasmoid-powered flares}
\label{sec:properties}
\subsection{Peak flare luminosity}
\label{sec:lum_size}
The observed peak luminosity of a flare  will depend on: (i) the plasmoid's size, $w^{\dpr}_{\rm f}$, when it leaves the reconnection layer, (ii) 
the plasmoid's Doppler factor at the end of Phase I, $\dpl(w^{\dpr}_{\rm f})$, (iii) the total number of radiating electrons, i.e., $N\propto n^{\dpr} \, w_{\rm f}^{\dpr 3}$, (iv) the timescale over which the plasmoid grows (dynamical) $t_{\rm dyn}\simeq w^{\dpr}_{\rm f}/c \vg$, and (v) the cooling timescale $t_{\rm cool}$ of radiating particles.  
Assuming that the radiating particles are fast cooling ($t_{\rm cool}\ll t_{\rm dyn}$), which is reasonable for UV/X-ray emitting electrons, the observed peak luminosity of a flare at a given frequency depends on (i)-(iv) and scales as 
\eqb 
\label{eq:Lpk_wf}
L_{\rm pk}(\nu) \propto \left[\dpl(w^{\dpr}_{\rm f})\right]^{3+\alpha} w^{\dpr 2}_{\rm f},
\eqe
with $\alpha=p/2$. The final size and momentum of a plasmoid carries information of its prior acceleration in the current sheet (see Fig.~\ref{fig:motion1}). Although larger plasmoids contain more particles, smaller plasmoids may power as luminous, or even more powerful, flares as their emission is more strongly Doppler boosted towards the observer. We argue that the brightest flares are not necessarily powered by the biggest plasmoids.

The above mentioned arguments are exemplified in Fig.~\ref{fig:Lpk_wf} where the measure of the peak luminosity, $[\dpl(w^{\dpr}_{\rm f})]^{3+\alpha} (w^{\dpr}_{\rm f}/\ell^\prime)^2$, is plotted against $w^{\dpr}_{\rm f}/\ell^\prime$ for different $\sigma$ (left to right). For each $\sigma$, the results for different final plasmoid sizes $w^{\dpr}_{\rm f}$ are shown. For all $\sigma$ values, the peak flare luminosity is not a monotonic function of the size for those plasmoids whose radiation is beamed towards the direction of the observer (here, for $\theta^\prime=\pi/4$ and $2 \pi/7$) and its dependence  on $w^{\dpr}_{\rm f}/\ell^\prime$ can be understood as follows. For favorable orientations between the layer and the observer the plasmoid's Doppler factor is  $\dpl \simeq 2 \Gamma_{\rm p} = 2\Gamma_{\rm j}\Gco(1+\beta_{\rm j}\bco\cos\theta^\prime)$. Plasmoids leave the layer with non-relativistic speeds if $w_{\rm f}^{\dpr}>w^{\dpr}_{\rm f, c}$, where 
\eqb
w^{\dpr}_{\rm f, c} \simeq 0.1 \, \beta_{\rm acc, -1} \ell^\prime.
\label{eq:wfc}
\eqe
In the above, we used eq.~(\ref{eq:PIC}), the condition $\bco\Gco=1$ and the approximation $\tanh(x)\approx x$.\footnote{This approximation, albeit useful,  underestimates the actual $w_{\rm f,c}^{\dpr}$ by a factor of two. We numerically found that $\bco\Gco\approx 2$ for plasmoids with $\wf$ given by eq.~(\ref{eq:wfc}). Equivalently, the corresponding Doppler factor is higher than $2\Gj$ and,  in particular,  $\delta_{\rm p, f}\sim 4\Gj$ instead of $2\Gj$. Regardless, the Doppler factor of plasmoids leaving the layer with non-relativistic speeds is independent of the plasmoid's size.} The Doppler factor of plasmoids leaving the layer with non-relativistic speeds is $\sim 2\Gj$ and therefore independent of the plasmoid's size. 
Thus, the peak flare luminosity depends quadratically on $w^{\dpr}_{\rm f}/\ell^\prime$ for large plasmoids. This is exactly illustrated in  Fig.~\ref{fig:Lpk_wf} (see blue and orange coloured lines for $w_{\rm f}^{\dpr} > 0.2 \ell^\prime$). 

Similarly, $\dpl(w^{\dpr}_{\rm f})$ is independent of $w^{\dpr}_{\rm f}$ for plasmoids that leave the 
layer with $\bco\Gco\rightarrow\sqrt{\sigma}$. The characteristic size can be estimated as
\eqb
w^{\dpr}_{\rm f, \sqrt{\sigma}}\simeq \frac{\vacc\ell^\prime}{\sqrt{\sigma}}=0.03 \frac{\beta_{\rm acc,-1} \ell^\prime}{\sqrt{\sigma_1}}.
\label{eq:wfasy}
\eqe 
Plasmoids with final sizes $w^{\dpr}_{\rm f}\approx (\vg/3\vacc)\,w^{\dpr}_{\rm f,  \sqrt{\sigma}} <w^{\dpr}_{\rm f, \sqrt{\sigma}} $ have been accelerated to their asymptotic momentum even before exiting the layer. The respective Doppler factor is then $\dpl \simeq 4 \Gamma_{\rm j} \sqrt{\sigma}$ for $w^{\dpr}_{\rm f}\lesssim w^{\dpr}_{\rm f, \sqrt{\sigma}}$ and the peak flare luminosity scales as $w^{\dpr 2}_{\rm f}$ (see also Fig.~\ref{fig:Lpk_wf}). For  $w^{\dpr}_{\rm f, \sqrt{\sigma}} < w^{\dpr}_{\rm f} < w^{\dpr}_{\rm f,c}$ the plasmoid's Lorentz factor is $1\ll \Gco <\sqrt{\sigma}$ and scales as $\Gco \propto 1/\wf$; here,  eq.~(\ref{eq:PIC}) and the linear approximation of $\tanh(x)$ were used. For favorable orientations $\dpl \propto \Gco$ and the peak luminosity measure scales as $\propto w_{\rm f}^{\dpr -1-\alpha}$. The scaling is more accurate for higher $\sigma$ where  
where the range of Lorentz factors between 1 and $\sqrt{\sigma}$ is wider (see right panel in Fig.~\ref{fig:Lpk_wf}).

The scalings of the peak luminosity measure for $w^{\dpr}\gtrsim w^{\dpr}_{\rm f,c}$ and $w^{\dpr} \lesssim w^{\dpr}_{\rm f,\sqrt{\sigma}}$ apply also to the non-favorable orientations, e.g. $\theta^\prime=0$ (black solid lines). However, for  the intermediate regime where the Doppler factor is $\dpl\propto 1/\Gco \propto w^{\dpr}_{\rm f}$, the  peak luminosity measure  is  $\propto w^{\dpr 5+\alpha}$. The strong dependence on $\wf$ is more evident in the high $\sigma$ cases (right panel in Fig.~\ref{fig:Lpk_wf}).

For plasmoids with favorable orientation and $w^{\dpr}_{\rm f,\sqrt{\sigma}}<w^{\dpr}_{\rm f} < w^{\dpr}_{\rm f, c}$  we find that brighter flares are powered by smaller and faster plasmoids  when exiting the current sheet (see also Fig.~\ref{fig:motion1}). The extent of this intermediate region of plasmoid sizes is larger for higher $\sigma$. For $w^{\dpr}_{\rm f} > w^{\dpr}_{\rm f, c}$, more luminous flares are powered by larger plasmoids that leave the current sheet with non-relativistic speeds (in the jet's rest frame).  A similar trend is found for flares powered by plasmoids with $w_{\rm f}^{\dpr} < w^{\dpr}_{\rm f,\sqrt{\sigma}}$. The effect of the layer's orientation on the peak  luminosity is negligible for  flares produced by monster plasmoids with $\wf \gtrsim 0.2\ell^\prime$. 

The results presented in Fig.~\ref{fig:Lpk_wf}  are obtained for a fixed viewing angle of the observer ($\thobs=0.5/\Gj$). The dependence of the peak luminosity on the orientation of the plasmoid relative to the observer is illustrated in Fig.~\ref{fig:map-boost} for $\sigma=10$ and two plasmoids: a small, fast-moving with $w^{\dpr}_{\rm f}=0.04 \ell^{\prime}$ (left panel) and a large, slow-moving (monster plasmoid) with $w^{\dpr}_{\rm f}=0.2 \ell^{\prime}$ (right panel). 
A wide range of values for the peak luminosity measure is found for smaller plasmoids due to their relativistic motion. As larger plasmoids leave the layer without becoming relativistic the angular dependence of the peak luminosity measure is less pronounced. An ``on-axis'' observer ($\thobs \ll$2.9\textdegree$/\Gamma_{j,1}$) would see bright flares powered by a small plasmoid (left panel) for angles $\theta^\prime$  ranging roughly from 30\textdegree \, to 100\textdegree. In contrast, an ``off-axis'' observer  ($\thobs \gg$ 2.9\textdegree$/\Gamma_{\rm j,1})$ would observe bright flares from plasmoids moving at $\theta^\prime > 90$\textdegree. Relatively bright flares may therefore be expected from AGN jets whose axis is misaligned with our line of sight. The intra-day TeV variability observed from M87 could be such an example \citep{giannios_10b}. The maximum possible peak luminosity measure for a smaller plasmoid is $\sim$ 3 times larger than for a rare, monster plasmoid, as indicated by the colour bars.  
Because of the relativistic motion of the 
smaller plasmoids the range of $\theta^\prime$ that favors the brightest flares is smaller compared to the larger plasmoids. Especially, for monster plasmoids and an on-axis observer almost all orientations $\theta^\prime\sim 0$\textdegree-180\textdegree \, would lead to flares of similar peak luminosity. 

\begin{figure*}
\centering
\includegraphics[width=0.33\textwidth]{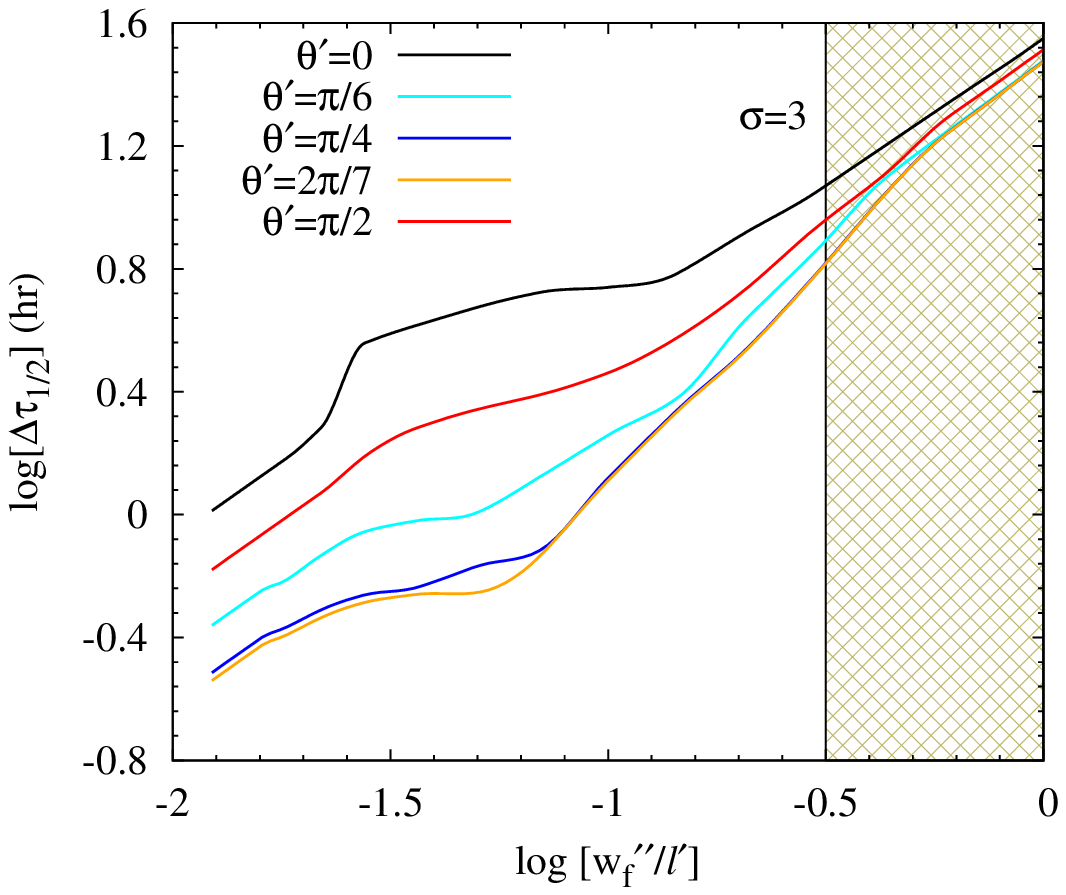}
 \includegraphics[width=0.33\textwidth]{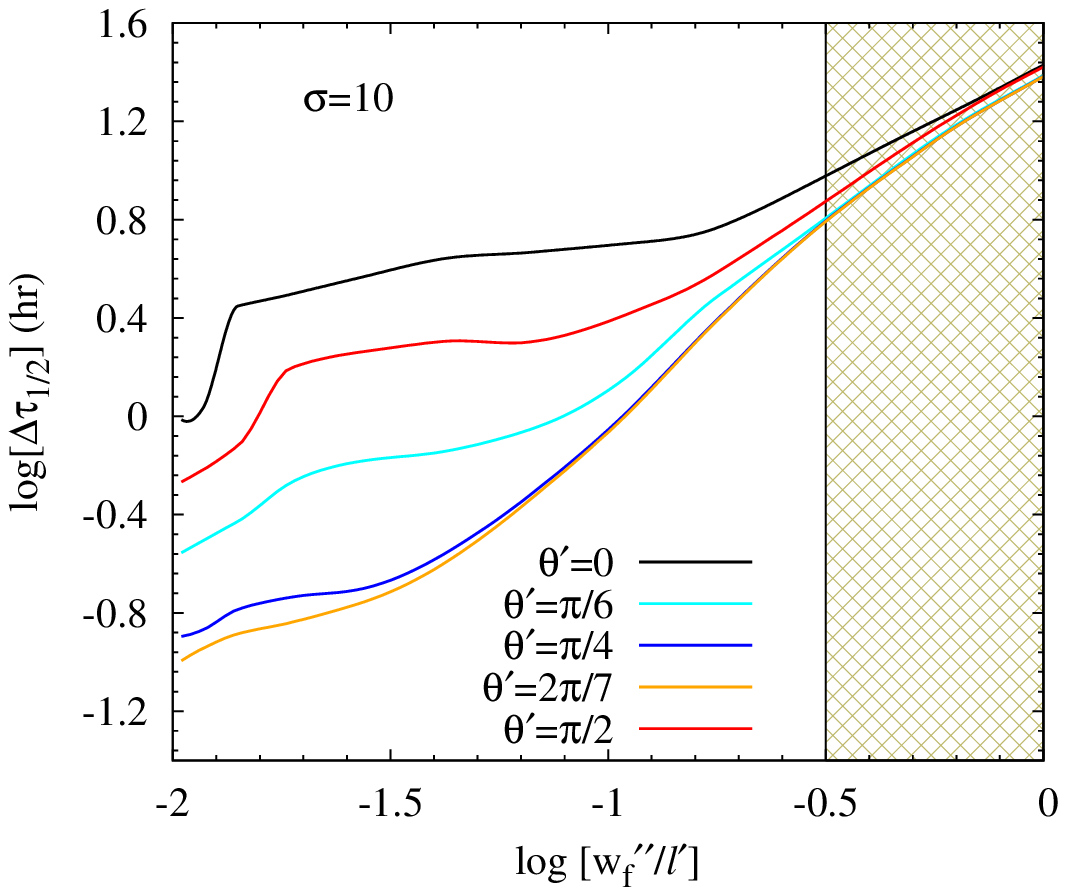}
 \includegraphics[width=0.33\textwidth]{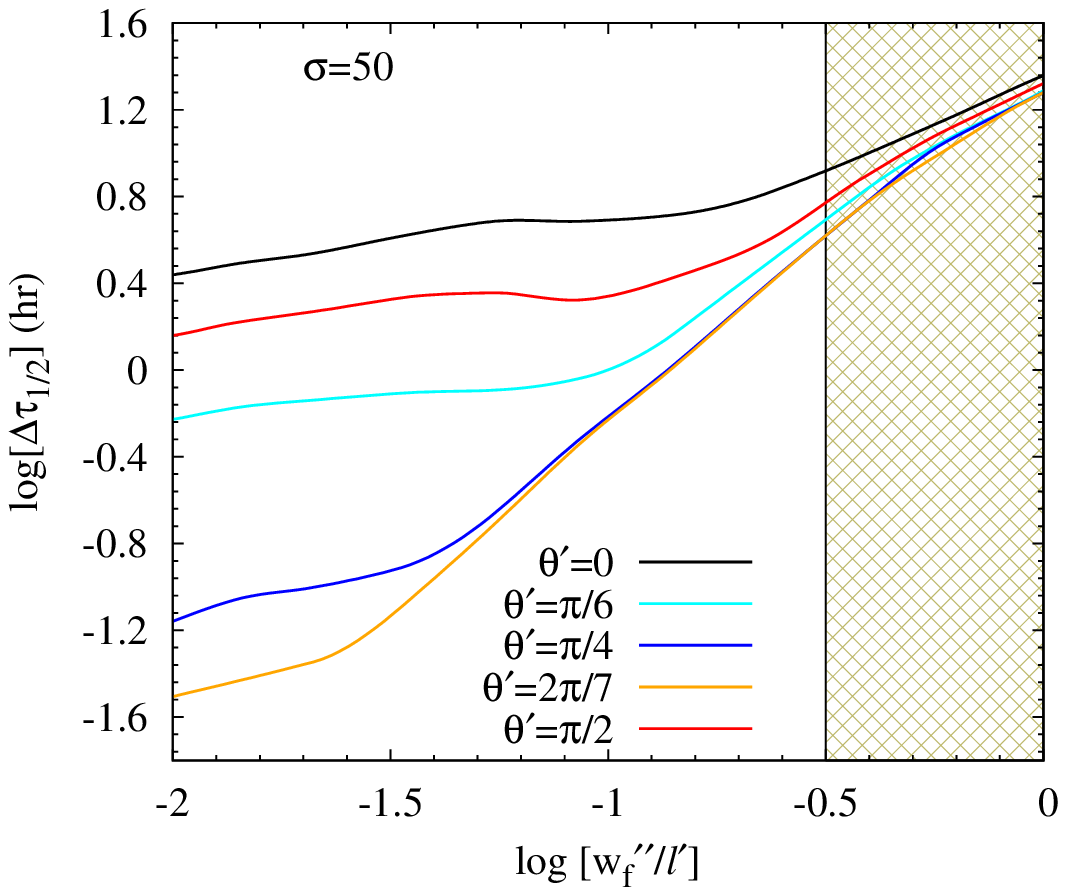}
 \caption{Log-log plot of the flux-doubling timescale $\Dtdb$, as determined by eq.~(\ref{eq:double}), versus $w^{\dpr}_{\rm f}/\ell^\prime$ (solid lines) for the same parameters as in Fig.~\ref{fig:motion1}, $\ell^\prime=10^{16}$~cm, and different values of $\theta^\prime$ marked on the plot. 
 The variability timescale does not depend much on $\theta^\prime$ for large pasmoids in contrast to smaller ones.} 
 \label{fig:Dt12_wf}
\end{figure*}

\subsection{Flux-doubling timescale}
\label{sec:timescale}
Among the different timescales characterizing a flare, the flux-doubling timescale is one that can be directly related
to the properties of the plasmoid powering the flare. This is defined as the time needed (in the observer's frame) for the luminosity (or, flux) to increase from half its peak value $L_{\rm pk}/2$ to its peak value $L_{\rm pk}$. 

As mentioned in the previous section, a plasmoid-powered flare reaches its peak luminosity when the plasmoid leaves the current sheet with a terminal size $w^{\dpr}_{\rm f}$ and Doppler factor $\dpl(w^{\dpr}_{\rm f})$. The size of the plasmoid, $w^{\dpr}_{1/2}$, at the moment the luminosity reaches half of its peak value is determined by the condition 
\eqb
\label{eq:condition}
\left[\dpl(w^{\dpr}_{1/2})\right]^{3+\alpha} w^{\dpr 2}_{1/2}=0.5 \left[\dpl(w^{\dpr}_{\rm f})\right]^{3+\alpha} w^{\dpr 2}_{\rm f},
\eqe
where we used eq.~(\ref{eq:Lpk_wf}). If the plasmoid's Doppler factor is constant, the above condition leads to 
$w_{1/2}^{\dpr}=\wf/\sqrt{2}$. This is a good approximation for plasmoids with $\wf \ll w_{\rm f, \sqrt{\sigma}}$ or for larger, non-relativistically moving plasmoids, i.e. $w_{\rm f}>w_{\rm f, c}$; in both cases, the plasmoid momentum is not significantly evolving over the relevant time window.
The flux-doubling timescale is then calculated using eq.~(\ref{eq:tobs}) as
\eqb
\label{eq:double}
\Dtdb (1+z)^{-1}= \int_{w^{\dpr}_{1/2}}^{w^{\dpr}_{\rm f}} \!\!\!{\rm d}\tilde{w}\frac{g(\tilde{w})}{c \vg \dpl(\tilde{w})}. 
\eqe
If the acceleration of the plasmoid and the suppression of its growth rate are ignored, the above expression results in
\eqb
\label{eq:estimate}
\Delta \tau_{1/2,\rm apr} (1+z)^{-1} \approx \frac{\wf}{\dpl(\wf) \vg c} \left(1-\frac{1}{\sqrt{2}}\right),
\eqe
which is a naive estimate of $\Dtdb$ as it underestimates the actual flux-doubling timescale (see Appendix \ref{app:naive}).

The flux-doubling timescale as determined by eq.~(\ref{eq:double}) for $\sigma =3, 10$ and 50 is illustrated in  Fig.~\ref{fig:Dt12_wf} for a fixed viewing angle $\thobs=0.5/\Gj$ and different angles $\theta^\prime$  marked on the plot.  
A wide range of flux-doubling timescales  is expected for plasmoid-powered flares. For the particular choice of $\ell^\prime=10^{16}$~cm, we find $\Dtdb$ from $\sim$ min to $\sim$ days. For small and  relativistically moving plasmoids the typical timescale may vary by two orders of magnitude among plasmoids with different $\theta^\prime$. In addition, shorter $\Dtdb$ are obtained for the higher $\sigma$ cases as a result of higher Doppler factors. For  $w^{\dpr}_{\rm f}>w^{\dpr}_{\rm f,c}$ (see eq.~(\ref{eq:wfc})), where both the Doppler factor and the growth rate are constant, the flux-doubling timescale scales linearly with respect to $w^{\dpr}_{\rm f}$ for all angles $\theta^\prime$. 

A similar trend is also expected for sufficiently small plasmoids that have been accelerated to $\beta_{\rm A}c$ before exiting the layer ($w^{\dpr}_{\rm f}\ll w^{\dpr}_{\rm f, \sqrt{\sigma}}$). The growth rate of such plasmoids is also constant but suppressed by a factor of three for our choice of the suppression factor $g(X^\prime/w^{\dpr})$. For the intermediate range of sizes $w^{\dpr}_{\rm f, \sqrt{\sigma}} \lesssim w^{\dpr}_{\rm f} < w_{\rm f, c}^{\dpr}$ the flux-doubling timescale depends sensitively on the acceleration and growth history of the plasmoids mainly through $\dpl$.  The exact shape of the curves in this intermediate range of sizes, which extends towards smaller $\wf$ for higher $\sigma$, since $w^{\dpr}_{\rm f,\sqrt{\sigma}}\propto \sigma^{-1/2}$, depends mainly on the plasmoid acceleration and angle $\theta^\prime$. The dependence on the latter is evident in Fig.~\ref{fig:Dt12_wf} for a given $\sigma$ value. Similarly, the curves for non-favorable orientations (e.g. $\theta^\prime=0$; 
black solid curves) differ among different magnetizations, as a result of 
differences in the plasmoid acceleration.

The flux-doubling timescale  of flares powered by plasmoids with favorable orientation (blue and orange lines) and $0.03< w^{\dpr}_{\rm f}/\ell^\prime < 0.3$ changes by at least one  order of magnitude, as illustrated in Fig.~\ref{fig:Dt12_wf}. On the contrary, the peak luminosity of such flares varies by a factor of a few only (see Fig.~\ref{fig:Lpk_wf}). For plasmoids whose radiation is beamed away from the observer (e.g. $\theta^\prime=0$; black solid curves) both the observed luminosity and flux-doubling timescales are  strongly dependent on $w^{\dpr}_{\rm f}$.

The two-dimensional maps of the flux-doubling timescale for $\sigma=10$ and two plasmoids with $w^{\dpr}_{\rm f}=0.04\ell^\prime$ (left) and $0.2\ell^\prime$ (right) are shown in Fig.~\ref{fig:map-doubling}. We find that the $\Dtdb$ of flares powered by larger plasmoids is approximately independent of $\theta^\prime$ for all viewing angles (right), whereas it strongly depends upon the combination $\theta^\prime$ and $\thobs$ for flares produced by smaller plasmoids (left). Inspection of Figs.~\ref{fig:map-boost} and \ref{fig:map-doubling} reveals a clear anti-correlation between the peak luminosity and flux-doubling timescale. Observing ultra-fast ($\lesssim 10-20$~min) and bright flares  (as e.g. in PKS 2155-304 \citep{aharonian_07} and Mrk 501 \citep{albert_07}),  requires favorable orientation of the current sheet towards the observer \citep[see also][]{giannios_09}. 


\begin{figure*}
\centering
\includegraphics[width=0.49\textwidth]{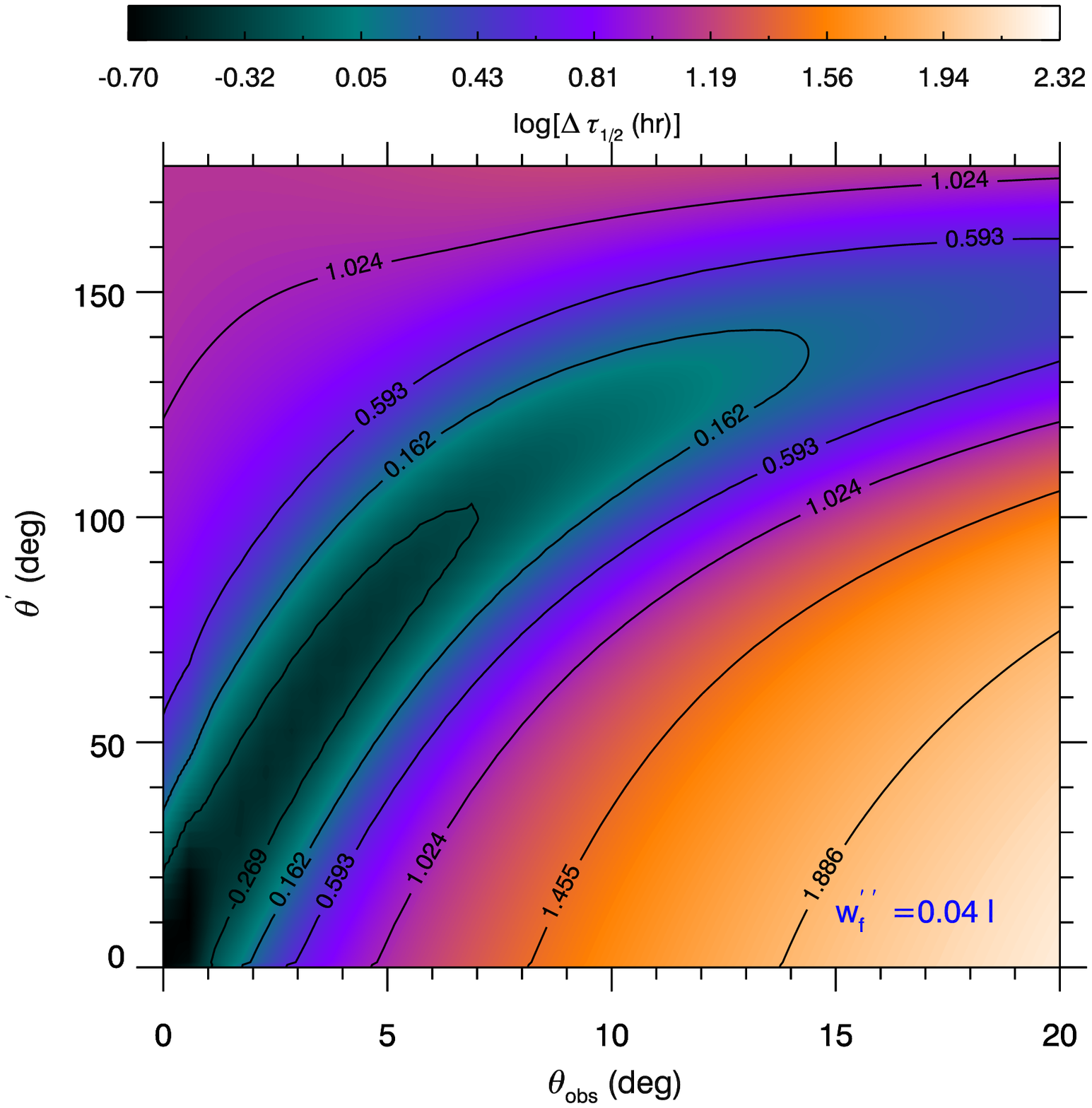}
\includegraphics[width=0.49\textwidth]{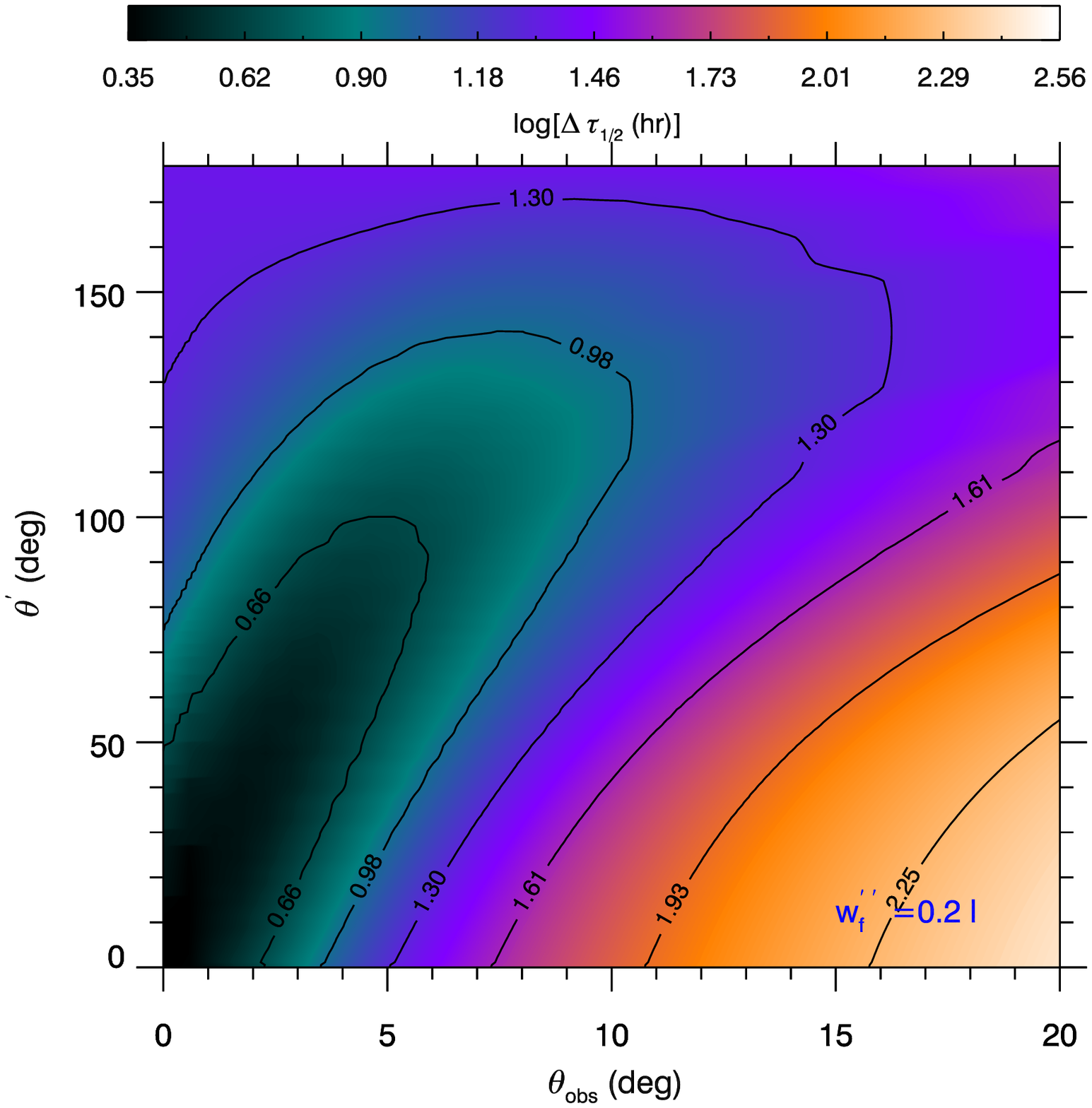}
 \caption{Two-dimensional map of the flux-doubling timescale for $\sigma=10$ and two plasmoids with $w^{\dpr}_{\rm f}=0.04\ell^\prime$ (left) and $0.2\ell^\prime$ (right). Each map is created for different viewing angles $\thobs$ and directions of plasmoid's motion with respect to the jet axis $\theta^\prime$ (in the jet's rest frame). Here, $\ell^\prime=10^{16}$~cm is assumed.}
\label{fig:map-doubling}
\end{figure*} 
\section{Application to the blazar emission}
\label{sec:results}
 Jets are likely to be launched as Poynting-flux dominated flows \citep[e.g.][]{blandford_znajek77, blandford_payne82} and remain such after their acceleration and collimation \citep[e.g.][]{spruit_96, vlahakis_04}. In this picture reconnection is a natural candidate for the dissipation of energy and emission (for reasons why shocks are disfavored, see also SPG15) provided that the magnetic field reverses polarity in the jet. Such reversals may result from MHD instabilities. Current driven instabilities, in particular, re-organize globally the jet's magnetic field and may lead to current sheets occupying a large fraction of the jet's cross section. However, the distance at which these instabilities may develop in  the jet and the resulting size of the current sheets remain poorly understood.  Alternatively, the jet could be a striped wind \citep[e.g.][]{lovelace_97, lyubarsky_kirk01, drenkhahn_02a,  mckinney_uzdensky12}. If the magnetic field reverses polarity  on a short timescale $\sim 10\, R_{\rm g}/c$ (
as, e.g., motivated by \citealt{parfrey_15}), where $R_{\rm g}\simeq 3\times 10^{14}\, M_9$~cm is the gravitational radius of a $M_9=M/10^9\, M_{\odot}$ black hole, the resulting  stripes would have width $\sim 3\times 10^{15}\,M_9$~cm (in the black hole frame) and the scale of the reconnection layers (in the jet frame)  would  be $\ell^\prime =3\times 10^{16}  \, \Gamma_{\rm j, 1} M_{9}$~cm. It can then be shown that  the dissipation takes place at a distance $z_{\rm diss}\simeq \Gj \ell^\prime /\beta_{\rm rec}\sim 1$ pc, where $\beta_{\rm rec} \sim 0.1$ is the reconnection speed \citep{giannios_13}. The location of the blazar zone is hotly debated and is proposed to lie either at sub-pc scales \citep[e.g.][]{ghisellini_tavecchio10, tavecchio_10, nalewajko_14} or at multi-pc scales \citep[e.g.][]{sikora_08, marscher_10,tavecchio_13,petro_dimi_15}. Regardless, the pc-scale location lies in middle of most estimates. Thus, we adopt $z_{\rm diss}=1$ pc and $\ell^\prime = 10^{16}$ cm as characteristic values.

The absolute power of a two-sided jet is written as \citep[e.g.][]{celotti_08, dermer_menon09}
\eqb
\label{eq:ljet}
L_{\rm j} = 2 \pi \varpi^2 c \beta_{\rm j}\Gamma_{\rm j}^2 (u^\prime_{\rm j}+P_j^\prime),
\eqe
where $\varpi \simeq z\theta_{\rm j}$ is the jet's cross section,  $\beta_j\approx 1$,  $u^\prime_{\rm j}$ and $P_j^\prime \sim u^\prime_j$ are the energy density and pressure of the unreconnected jet flow.  The energy density in the jet's frame is given by
 \eqb 
\label{eq:uj}
u^\prime_{\rm j} \simeq  2.6\, L_{\rm j, 46} \varpi^{-2}_{16} \Gamma_{\rm j, 1}^{-2}\, {\rm erg \, cm^{-3}}.
\eqe 
and is approximately equal to the energy density in the plasmoid's rest frame, i.e. $u^{\dpr}_{\rm j}\sim u^\prime_{\rm j}$ (SGP16).
The blazar jet composition is uncertain but several observations \citep[e.g.][]{celotti_08, ghisellini_10, ghisellini_14} are compatible with baryon-loaded jets with several pairs per proton. Since relativistic reconnection generally results in the same energy per particle regardless of the particle species, we neglect the proton contribution to the total energy density of a plasmoid, which is taken to be (moderately) dominated by leptons over the magnetic field (SPG15).  This translates to $u^{\dpr}_{\rm j} \sim u^{\dpr}_{\rm e} \gtrsim u^{\dpr}_{\rm B}$. 
\subsection{Observables}
\label{sec:observables}
Here we derive simple analytical expressions for two observables of flares, namely their peak luminosity $L_{\rm pk}$ and 
flux-doubling timescale $\Dtdb$. We neglect the evolution of $\vg$ with the plasmoid size (see \S\ref{sec:model}), we assume perfect alignment between the layer and the observer and consider flares produced by two types of plasmoids: (i) those that exit the layer with $\bco\Gco=1$ and size $w_{\rm f,c}\sim 0.1 \beta_{\rm acc,-1}\ell^\prime$ (see eq.~(\ref{eq:wfc})) and those with $\bco\Gco=\sqrt{\sigma}$ and size $w_{\rm f, \sqrt{\sigma}}=0.03 \beta_{\rm acc,-1} \ell^\prime/\sqrt{\sigma_1}$; for the definition see eq.~(\ref{eq:wfasy}). In general, the bolometric peak flare luminosity can be written as 
\eqb
\label{eq:Lbol}
L_{\rm pk,bol}=\frac{\pi}{2} \vg c w_{\rm f}^2 \delta_{\rm p,f}^4 u^{\dpr}_{\rm e}
\eqe
where $u^{\dpr}_{\rm e}\simeq f_{\rm rec} u^\prime_{\rm j} = f_{\rm rec} L_{\rm j}/ 4\pi \varpi^2 c \beta_{\rm j} \Gamma_{\rm j}^2$. The expression for the peak bolometric luminosity is derived under the assumption that the radiating particles are fast cooling ($t_{\rm cool}\ll t_{\rm dyn}$). This is, in general, true for monster plasmoids (see \S\ref{sec:examples}), but for smaller plasmoids the expressions for the peak bolometric luminosities should be treated as upper limits.
\begin{enumerate}
 \item Plasmoids with $w_{\rm f,c}$: substitution of $w_{\rm f,c}$ in the above expression results in
\eqb
\label{eq:Lpk-nr}
L_{\rm pk,bol}= 32\, f_{\rm rec} L_{\rm j}\Gamma_{\rm j}^2  \vg \vacc^2 \beta_{\rm rec}^2 \left(\Gamma_{\rm j}\theta_{\rm j}\right)^{-2},
\eqe
where we also used $\ell^{\prime}/\varpi\approx \beta_{\rm rec}/\Gamma_{\rm j}\theta_{\rm j}$ and $\delta_{\rm p,f}\approx 4\Gamma_{\rm j}$ (see footnote following  eq.~(\ref{eq:wfc})). The flux-doubling timescale can be  calculated using eq.~(\ref{eq:double}) and is given by 
\eqb
\label{eq:double-nr}
\Dtdb \simeq \frac{\vacc \ell^\prime}{4\Gamma_{\rm j}\vg c} \simeq 2.3\,\frac{\beta_{\rm acc,-1}\ell^\prime_{16}}{\beta_{\rm g,-1}\Gamma_{\rm j, 1}}\,{\rm hr}.
\eqe
The energy release during $\Dtdb$ can be approximated by
\eqb
\label{eq:Ebol-nr}
E_{\rm bol} \approx L_{\rm pk, bol} \Dtdb \simeq \frac{8}{c}f_{\rm rec}L_{\rm j}\Gamma_{\rm j} \left(\Gamma_{\rm j} \theta_{\rm j}\right)^{-2}\vacc^3\beta_{\rm rec}^2 \ell^\prime.
\eqe
Interestingly, the energy release during flares powered by monster plasmoids ($w_{\rm f,c}=0.1\beta_{\rm acc,-1}\ell^\prime$) does not depend strongly on the
magnetization of the jet, since $\vacc$, $\beta_{\rm rec}$, and $f_{\rm rec}$ are almost independent of $\sigma$, as shown in SPG15 and SGP16. It is the large-scale properties of the jet (e.g. $\Gamma_{\rm j}, L_{\rm j}$) and the length of the reconnecting layer that determine the fluence of such flares. 
\item Plasmoids with $w_{\rm f,\sqrt{\sigma}}$: substitution of $w_{\rm f,\sqrt{\sigma}}$ in eq.~(\ref{eq:Lbol}) leads to  
\eqb
\label{eq:Lpk-asy}
L_{\rm pk,bol}= 32 \, f_{\rm rec} L_{\rm j}\Gamma_{\rm j}^2 \sigma \beta_{\rm g}^{\rm as} \vacc^2 \beta_{\rm rec}^2 \left(\Gamma_{\rm j}\theta_{\rm j}\right)^{-2},
\eqe
where $\beta_{\rm g}^{\rm as}\simeq \vg/3$ (the growth of small plasmoids is suppressed by a factor of 3, for our choice of $g(X^\prime/w^{\dpr})$) and $\vg$ listed in Table~\ref{tab1}. This is to be compared to eq.~(\ref{eq:Lpk-nr}) where $\vg$ appears, since the growth rate is not suppressed for non-relativistic plasmoids. Thus, the bolometric peak luminosity of flares produced by small and fast moving plasmoids is higher by a factor of $\sigma(\beta_{\rm g}^{\rm as}/\vg)\simeq \sigma/3$ compared to $L_{\rm pk, bol}$ of flares  produced by the larger and non-relativistic plasmoids.  The flux-doubling timescale is given by
\eqb
\label{eq:double-asy}
\Dtdb \simeq \frac{\ell^\prime \vacc}{4\Gamma_{\rm j}\sigma \beta^{\rm as}_{\rm g} c} \simeq 14\, \frac{\ell^\prime_{16} \beta_{\rm acc,-1}}{\sigma_1\Gamma_{\rm j,1} \beta^{\rm as}_{\rm g,-1}}\, {\rm min},
\eqe
where we used $\delta_{\rm p,f}\approx 4\Gamma_{\rm j}\Gco\approx 4\sqrt{\sigma}\Gamma_{\rm j}$. The fluence of a flare over $\Dtdb$ is written as
\eqb
\label{eq:Ebol-asy}
E_{\rm bol} \approx L_{\rm pk, bol} \Dtdb \simeq \frac{8}{c}f_{\rm rec}L_{\rm j}\Gamma_{\rm j} \left(\Gamma_{\rm j} \theta_{\rm j}\right)^{-2}\vacc^3\beta_{\rm rec}^2 \ell^\prime,
\eqe
which is same as eq.~(\ref{eq:Ebol-nr}). Thus, the observed fluence of flares powered by either monster plasmoids or much smaller plasmoids with $\bco \Gco \rightarrow \sqrt{\sigma}$ that form in the same reconnecting layer is nearly the same despite of the large differences in $\Dtdb$ and $L_{\rm pk}$. Interestingly, the $E_{\rm bol}$ is nearly independent of the plasma magnetization, since $\vacc$ and $\beta_{\rm rec}$ do not vary much with $\sigma$. 

Observations of ultra-fast ($\sim$min) and slower varying ($\sim$ hr) flares of similar fluence during a longer period of blazar flaring activity (e.g. lasting for several hours to days) would indicate the emission from the same current sheet. The ratio of the flux-doubling timescales of these flares is $\simeq 2\sigma$ (see eq.~(\ref{eq:double-nr}) and (\ref{eq:double-asy})), i.e. it can be used to infer the magnetization on the jet. 
The inferred $\sigma$ could be potentially cross-checked with the inferred slope of the particle distribution, which also depends on the magnetization (see \S\ref{sec:examples}). For favorable orientations, both fast and slower varying flares are expected to have similar peak flare luminosities, so they are likely to be both detectable.
\end{enumerate}

The observed timescale required for the energy to be dissipated in an active region of the jet with (comoving) volume $\sim \ell^{\prime 3}$ defines the total observed duration of a single reconnection event and is given by  $\sim \ell^\prime/\Gamma_{\rm j} \beta_{\rm rec} c \simeq 88\, {\rm hr}\, \ell^{\prime}_{16} \Gamma^{-1}_{\rm j, 1}$. The dissipated energy will be radiated away in multiple bursts over this longer period. The largest (monster) plasmoids leave the layer at non-relativistic or mildly-relativistic speeds giving rise to flares of duration $\lesssim \ell^\prime/\delta_{\rm j} c\simeq 8 \, {\rm hr} \, \ell^\prime_{16} \delta^{-1}_{\rm j, 1}$ (see e.g. eq.~(\ref{eq:double-nr}) and Fig.~\ref{fig:Dt12_wf}). Several  flares with such durations may be expected from a single reconnection event, since monster plasmoids are ejected from the current sheet every few dynamical times (see also SGP16). Smaller plasmoids may produce much faster flares depending on the orientation of the observer and the 
reconnecting layer. For favorable orientations, several bright and ultra-fast ($\sim 10-20$ min) flares may be produced by smaller plasmoids that are frequently formed in the reconnection layer. 

Detection of multiple flares within a period of increased activity could serve as a  probe} of the layer's orientation with respect to the observer. This is exemplified in Fig.~\ref{fig:Lpk_Dtdb} where the peak  (bolometric) luminosity of flares produced by plasmoids of different sizes is plotted against the respective flux-doubling timescale. The results are shown for $\sigma=10$,  $f_{\rm rec}=0.5$, $\thobs=0.5/\Gamma_{\rm j}$, and various angles $\theta^\prime$.  The growth rate is assumed to be constant and equal to $\vg/(1+2\tanh(2\beta_{\rm co,f}/\beta_{\rm A})$, where $\beta_{\rm co,f}$ is the final plasmoid velocity. Here, the jet parameters are $\ell^\prime=10^{16}$~cm, $L_{\rm j}=10^{46}$~erg s$^{-1}$, $\Gamma_{\rm j}=10$, and $\Gamma_{\rm j}\theta_{\rm j}=0.2$ \citep{pushkarev_09}. Similar results are expected for other magnetizations except for the minimum $\Dtdb$ which will be lower for higher $\sigma$.  Data from flares with a range of observed rise timescales could be included 
in Fig.~\ref{fig:Lpk_Dtdb} in order to infer $\theta^\prime$ for an on-axis observer.

\begin{figure}
 \centering
 \includegraphics[width=0.48\textwidth]{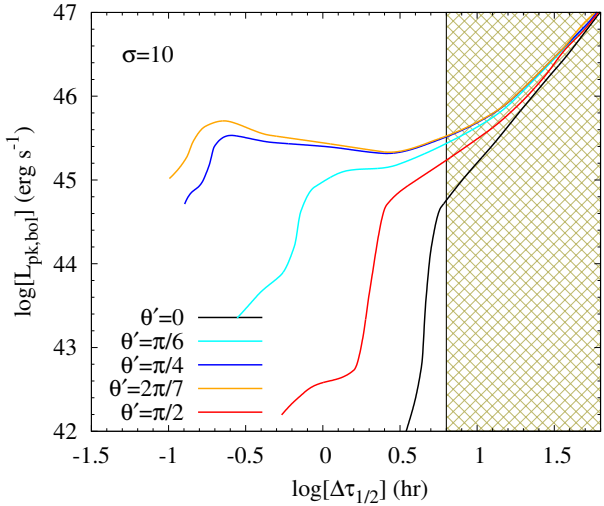}
 \caption{Bolometric peak luminosity  of flares produced by plasmoids of different sizes as a function of the respective flux-doubling timescale. Coloured curves correspond to different orientations of the reconnecting layer with respect to the jet axis (see inset legend) with the optimal orientation obtained for $\theta^\prime=\pi/4\dots2\pi/7$. Flares characterized by $\Dtdb \gtrsim 6.3$~hr are produced by extremely rare plasmoids with $\wf\gtrsim 3\times10^{15}\ell^\prime_{16}$~cm (see also Fig.\ref{fig:Dt12_wf})). Other parameters used are:  $\Gamma_{\rm j}=10$,  $\thobs=0.5/\Gamma_{\rm j}$, $L_{\rm j}=2\times 10^{46}$~erg s$^{-1}$, $f_{\rm rec}=0.5$, and $\Gamma_{\rm j}\theta_{\rm j}=0.2$.}
 \label{fig:Lpk_Dtdb}
\end{figure}

\subsection{Indicative examples}
\label{sec:examples}
Henceforth, we adopt $u^{\dpr}_{\rm e}\simeq 3 u^{\dpr}_{\rm B}$ as inferred from PIC simulations for all $\sigma \ge 3$ (see also SPG15)  and, as an example, we set $u^{\dpr}_{\rm e}=0.5$ erg~cm$^{-3}$. This translates  to $u^{\dpr}_{\rm B}=0.15$ erg~cm$^{-3}$ or $\Bp\simeq2$~G and {corresponds to $L_{\rm j}=2\times10^{46}$~erg s$^{-1}$ for $f_{\rm rec}=0.5$, $\beta_{\rm rec}=0.1$, $\Gj=10$, $\Gj\theta_{\rm j}=0.2$, and $\ell^\prime=10^{16}$~cm. The adopted energy densities} are within the ranges inferred from leptonic modelling of blazar emission \citep[e.g.][]{celotti_08,tavecchio_16} and will be used throughout this section as reference values. In all cases presented below the plasmoid is assumed to be formed at  $X^\prime_0=10^{-2}\ell^\prime$; we also considered $\theta^\prime=\pi/4$ and $\thobs=0.5/\Gj$. For each of the different magnetizations considered here (see Table~\ref{tab1}) we present indicative cases of a small/relativistic plasmoid and large/non-relativistic plasmoid (for the parameters, 
see Table~\ref{tab2}).

\begin{table}
\centering
\caption{Plasmoid growth and acceleration rates for different magnetizations. The characteristics of the particle distribution at injection are also listed. The expansion velocity in Phase II is {assumed to be constant ($\alpha=0$) and }fixed to $\beta_{\rm exp}c=0.08c$ in all cases. {Among the listed quantities only $\beta_{\rm exp}$ and the pair multiplicity $N_{\pm}$ are free parameters. All other quantities can be benchmarked with PIC simulations of reconnection.}}
\begin{threeparttable}
   \begin{tabular}{c ccc}
  \hline 
$\sigma$ & 3 & 10 & 50 \\
$\vacc$ & 0.12 & 0.12 & 0.15 \\
$\vg$\tnote{\textdagger} & 0.06 & 0.08 & 0.1 \\
$f_{\rm rec}$\tnote{\textdaggerdbl}& 0.15 & 0.25 & 0.5 \\
$p$ & 3 & 2.1 & 1.5 \\
$\gamma_{\min}$\tnote{*} & 140 & 200 & 1 \\
$\gamma_{\max}$\tnote{*} &  $2\times 10^4$ & $2\times 10^4$ & $2.5\times 10^3$ \\
$N_{\pm}$ &  1.5   & 1.0  & 450\\
\hline
 \end{tabular}
  \tnote{\textdagger} The values refer to the growth rate without the suppression factor. \\ 
  \tnote{\textdaggerdbl} The values in the first two columns are appropriate for reconnection in electron-proton plasma with $N_{\pm}\sim 1$, whereas the third value is appropriate for electron-positron plasma; in fact, $N_{\pm} \gg 1$ in this case. \\
  \tnote{*} The values are derived using eqs.~(\ref{eq:gmin}) and (\ref{eq:gmax}).
  \
  \end{threeparttable} 
\label{tab1}
\end{table}
\begin{table}
\centering
\caption{Plasmoid size, momentum and Doppler factor used in the indicative cases for $\sigma=10$ presented in \S\ref{sec:lc-sed}.}
 \begin{tabular}{c c c }
  \hline 
  & Small \&  & Large (monster) \& \\
  & relativistic plasmoid &  non-relativistic plasmoid\\
    \hline
\hline
$w^{\dpr}_0$ (cm) & $3.2\times 10^{12}$ & $6\times 10^{12}$\\
$\wf$ (cm) & $4\times 10^{14}$  & $2\times 10^{15}$\\
$w^{\dpr}_{\rm f}/\ell^\prime$ & {0.04} & 0.2 \\
$\beta_{\rm co}\Gamma_{\rm co}$ &  2.4 & 0.6\\
$\delta_{\rm p, f}$ & {70.5} &  {29} \\
\hline
 \end{tabular}
 \label{tab2}
\end{table}

In the following, we present the multi-wavelength spectra and light curves obtained for indicative cases  after 
solving numerically the kinetic equations for electrons and photons. This allows us to include 
synchrotron self-Compton cooling as an additional energy loss process for electrons and calculate the synchrotron and SSC photon spectra. 
In addition, 
by solving numerically the kinetic equations we are able to investigate different forms of the injection rate profile in Phase II (a phase which cannot be benchmarked with PIC simulations) and their effects on the decay slope of the light curves.
\subsubsection{Numerical code}
The kinetic equations for electrons and photons written in the plasmoid's rest frame are given by
\eqb
\label{eq:num-elec}
\frac{\partial n^{\dpr}_{\rm e}}{\partial w^{\dpr}} + 3\frac{n^{\dpr}_{\rm e}}{w^{\dpr}} + \mathcal{L}_{\rm e}^{(\rm syn)}+\mathcal{L}_{\rm e}^{(\rm ics)}+\mathcal{L}_{\rm e}^{(\rm ad)} & = & \mathcal{Q}_{\rm e}^{\rm (inj)}+\mathcal{Q}_{\rm e}^{(\gamma \gamma)} \\
\frac{\partial n^{\dpr}_{\gamma}}{\partial w^{\dpr}} + 3\frac{n^{\dpr}_{\gamma}}{w^{\dpr}} +2 \frac{n^{\dpr}_{\gamma}}{\beta w^{\dpr}} + \mathcal{L}_{\gamma}^{(\gamma \gamma)} + \mathcal{L}_{\gamma}^{\rm (ssa)} & = & \mathcal{Q}_{\gamma}^{\rm (syn)} + \mathcal{Q}_{\gamma}^{\rm (ics)}
\label{eq:num-phot}
\eqe
where the second term in the left hand side of eqs.~(\ref{eq:num-elec}) and (\ref{eq:num-phot}) accounts for the dilution of the number density due to the increase of the volume.  The third term in the left hand side of eq.~(\ref{eq:num-phot}) describes the photon escape within a crossing time $w^{\dpr}/2c$ and $\beta=\vg$ or $\vexp$ for Phases I and II, respectively. The operators $\mathcal{L}$ denote particle energy losses and/or sinks of particles, while the operators $\mathcal{Q}$ denote terms of energy and/or particle injection. The physical processes which are included in the aforementioned equations are: (i) electron synchrotron radiation (``syn'') and synchrotron self-absorption (``ssa''); (ii) inverse Compton scattering (``ics''); (iii) photon-photon pair production (``$\gamma \gamma$'') and (iv) adiabatic losses (``ad''), which become relevant only in Phase II; in the first phase the plasmoid size increases due to the accumulation of fresh particles rather than the work done by the particles 
themselves (for more details, see \citet{petro_mast09}).

The particle injection rate in Phase I is benchmarked with PIC simulations and is modelled as $\mathcal{Q}_{\rm e, I}^{\rm (inj)}\propto w^{\dpr 2}$. This choice results in a constant particle density in the plasmoid. For the injection rate in Phase II, which cannot be constrained by PIC simulations, we adopt the following:
\eqb 
\label{eq:Q2}
\mathcal{Q}_{\rm e, II}^{\rm (inj)} \propto w^{\dpr 2}\exp\left[-\eta (w^{\dpr}-\wf)/\wf\right]
\eqe 
where $\eta$ is a free parameter. For $\eta \gg 1$, the above expression simulates a fast cessation of the electron injection rate with a characteristic timescale (in the observer frame) $t_{\rm off}\sim \wf/\eta \vg c \delta_{\rm p, f} \ll \wf/ c \delta_{\rm p, f}= t_{\rm cr,f}$; here,  $t_{\rm cr,f}$ is the light crossing time of the plasmoid (in the observer frame) at the end of Phase I. In practice, no fresh particles are injected for $\eta\gg1$. Thus, our results are independent of the modeling details of Phase II, such as the expansion profile of the plasmoid. In this regard, the results obtained for  $\eta \gg 1$ (henceforth, we adopt this value as our default case) are the most robust.
However, we also explored more slowly decaying injection rates for $\eta=$3,6, and 9.  Our default choice for the magnetic field decay in Phase II is $B^{\dpr}_{\rm p}\propto w^{\dpr -q}$ with $q=1$. In addition, we considered a faster decaying magnetic field in Phase II described by $q=2$. 

\subsubsection{Light curves and spectra}
\label{sec:lc-sed}
 \begin{figure*}
\includegraphics[width=0.48\textwidth]{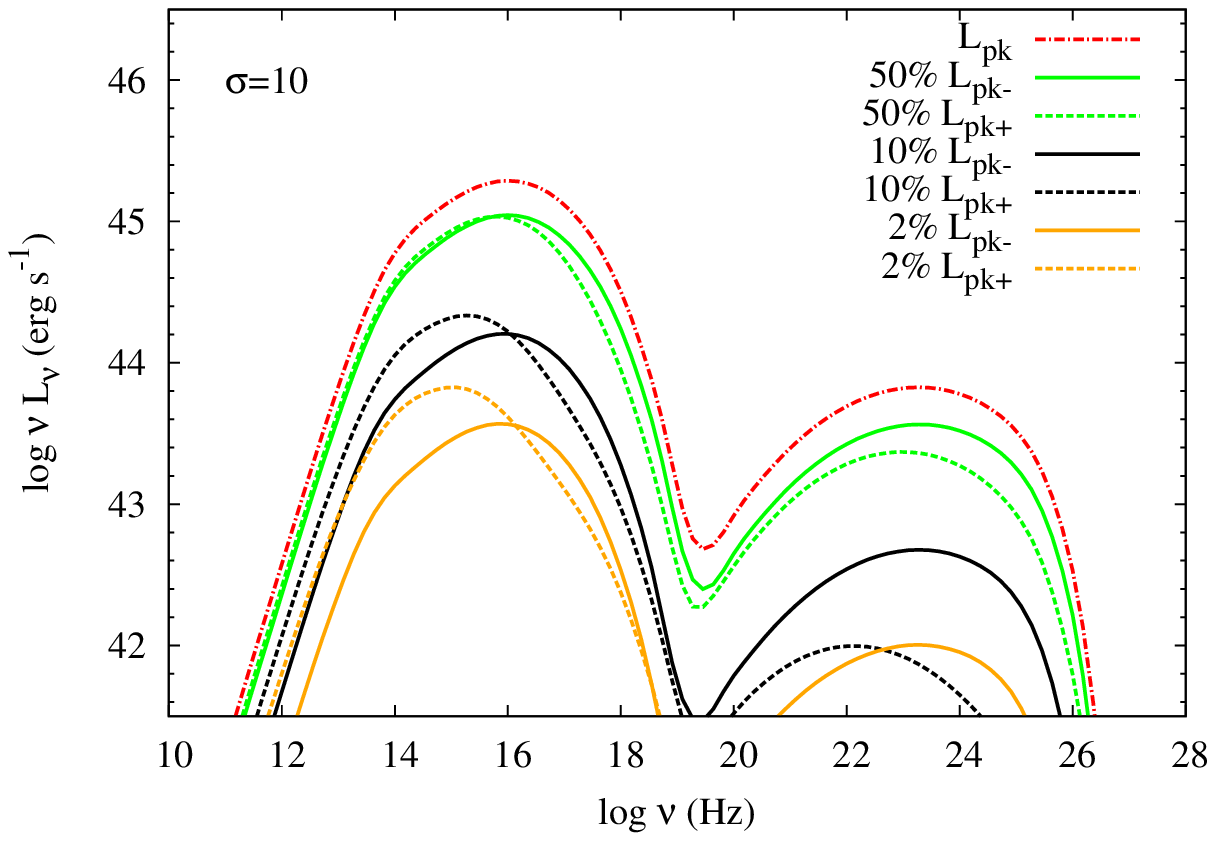}
\includegraphics[width=0.48\textwidth]{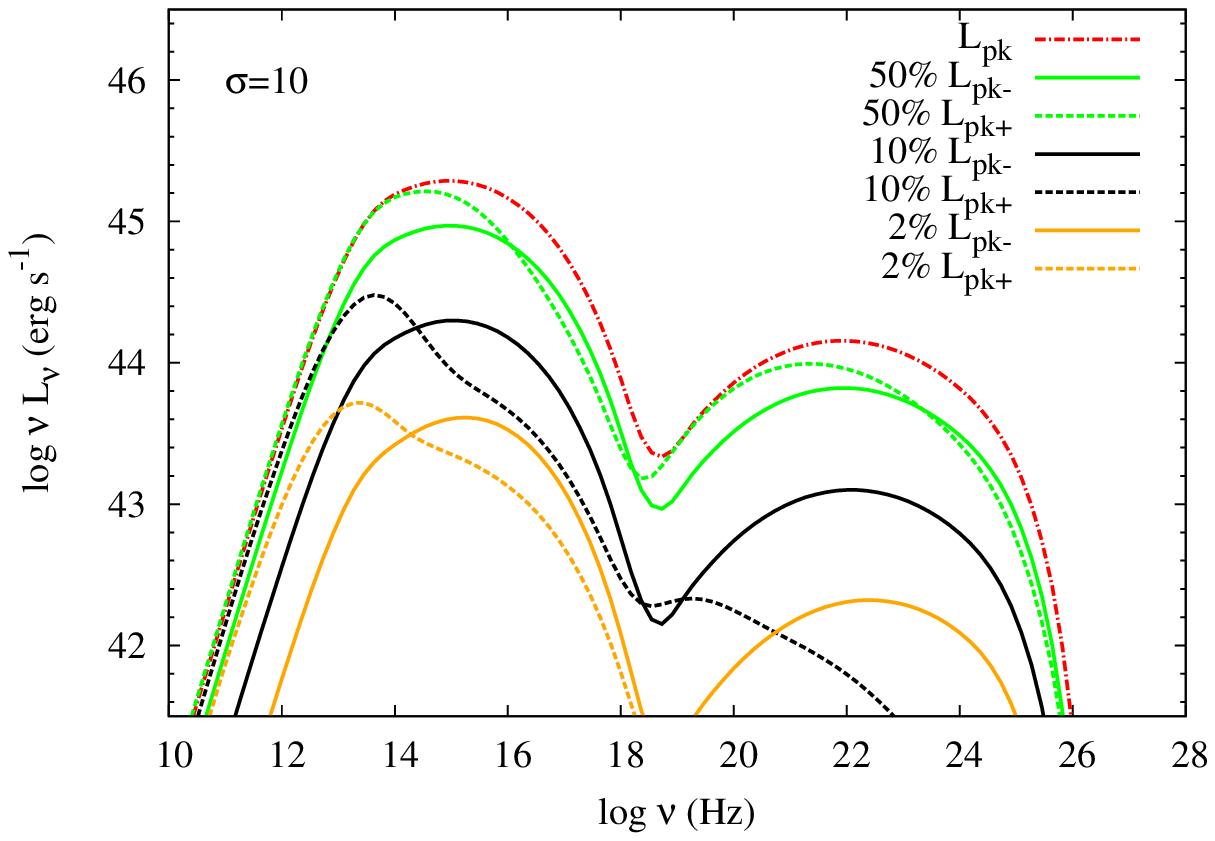}
\caption{Observed multi-wavelength spectra of a flare produced by a small relativistic plasmoid (left panel) and a large (monster)  non-relativistic plasmoid (right panel). The spectra obtained at the peak time of the flare are plotted with red dot-dashed lines. Other snapshots corresponding to times where the luminosity reaches $X\%$ of the peak bolometric luminosity are overplotted with solid (before the peak) and dashed (after the peak) lines. Here, $\sigma=10$.  All other parameters are listed in Tables~\ref{tab1} and \ref{tab2}.}
\label{fig:fig1}
\end{figure*}
\begin{figure*}
\includegraphics[width=0.48\textwidth]{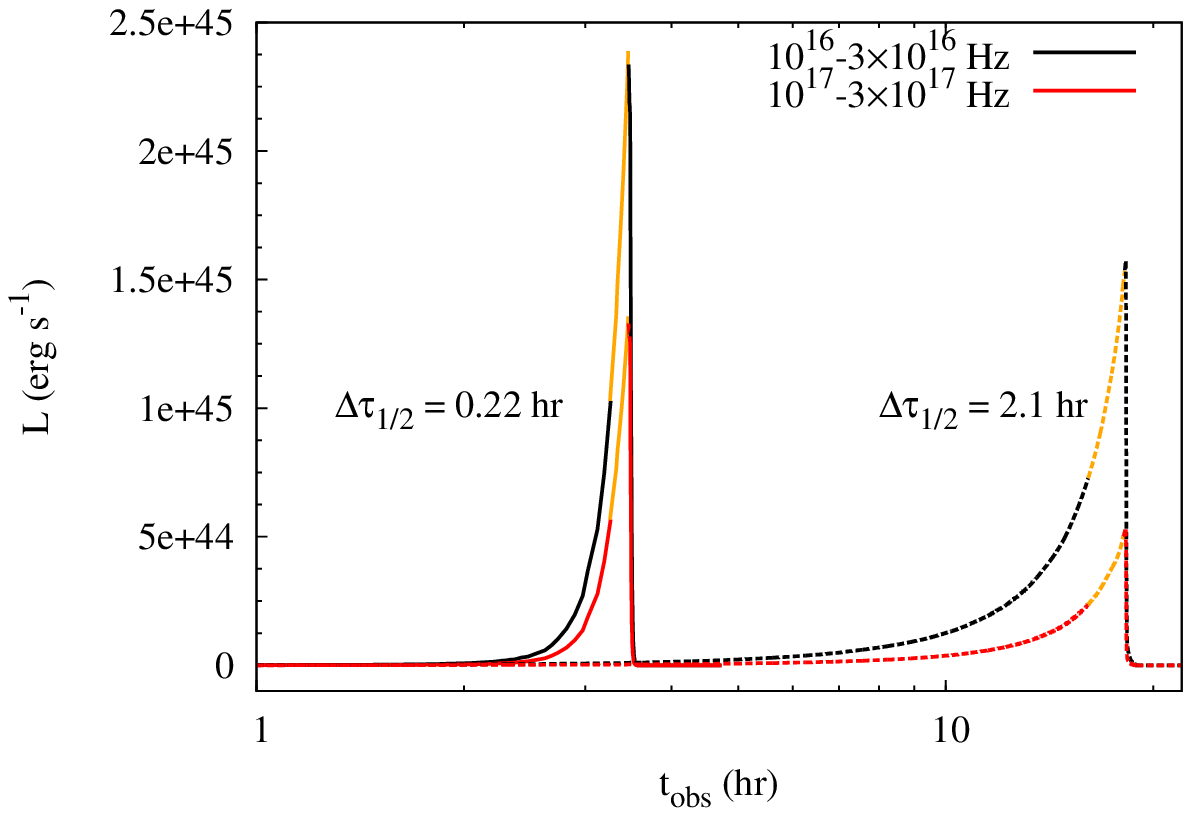}
\includegraphics[width=0.48\textwidth]{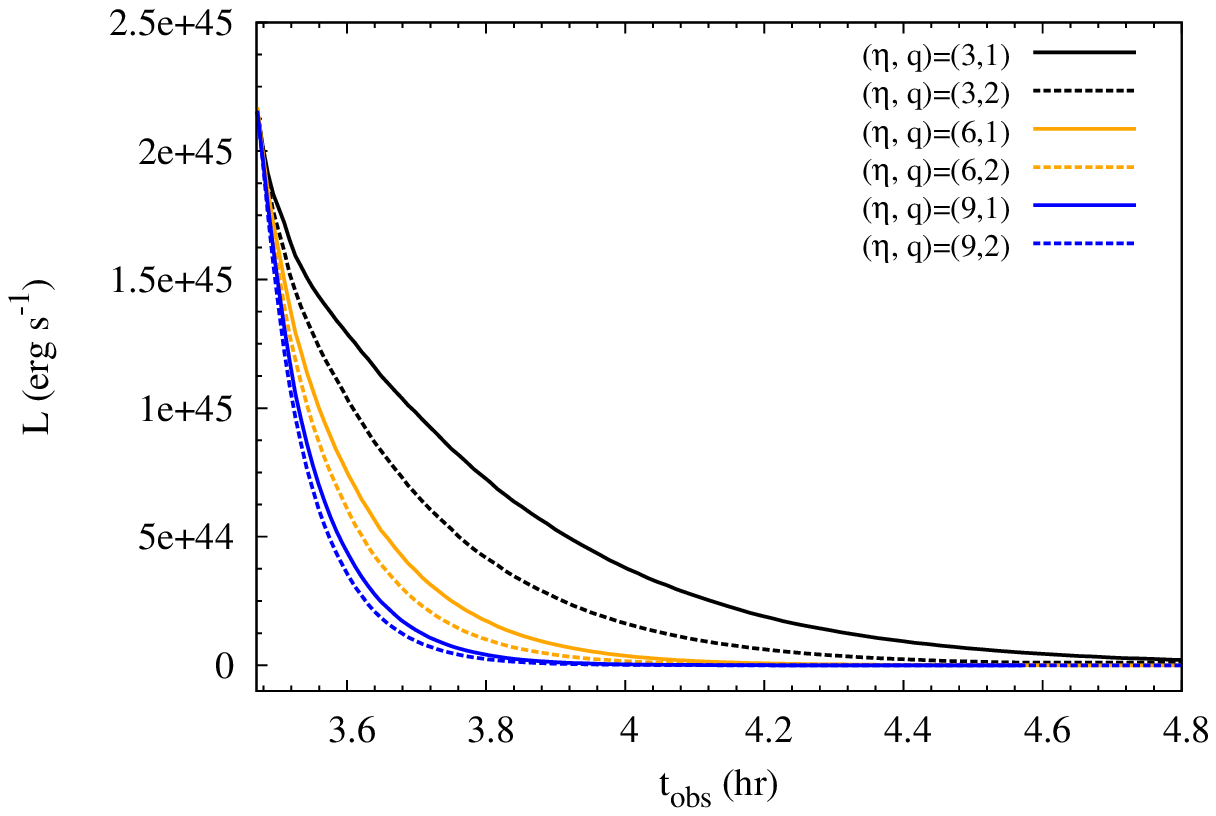}
\caption{Left panel: Light curves at the frequency bands $\nu=10^{16}-3\times10^{16}$~Hz (black lines) and $\nu=10^{17}-3\times10^{17}$ Hz (red lines) obtained for the default case of a fast cessation of particles at the end of Phase I. The light curves with the shortest duration and flux-doubling timescale $\Dtdb=0.22$~hr (solid lines) correspond to a flare produced by a small and relativistic plasmoid, whereas those obtained for a monster plasmoid have  $\Dtdb \sim 2.1$~hr (dashed lines). In both cases, the flux-doubling timescale is in agreement with Fig.~\ref{fig:Dt12_wf} (middle panel). The part of the light curves that corresponds to the flux-doubling timescale is highlighted with orange colour. The properties of the depicted light curves are benchmarked with PIC simulations.  Right panel: 
Zoom in the decaying part of slower decaying flares than those shown in the left panel. Here, $\nu=10^{16}-3\times10^{16}$~Hz, $\sigma=10$ and different values of $\eta$ and $q$ marked on the plot (for details, see text). The results are obtained for a small and 
relativistic plasmoid (see Table~\ref{tab2}).}
\label{fig:fig2}
\end{figure*}

The multi-wavelength photon spectra produced by a small relativistic plasmoid and a monster plasmoid for $\sigma=10$ are presented in the left and right panels, respectively, of Fig.~\ref{fig:fig1}.  The snapshots are obtained at the peak time of the flare (red dot-dashed line), at the rising part (solid lines) and at the decaying part (dashed lines) of the flare. Spectra that correspond to $X\%$ of the peak bolometric luminosity are shown with different colours. The resulting light curves at two frequency bands ($10^{16}-3\times10^{16}$~Hz and $10^{17}-3\times10^{17}$~Hz) are shown in Fig.~\ref{fig:fig2} (left panel). The results are obtained for the default choice of the injection rate in Phase II given by eq.~(\ref{eq:Q2}) and $\eta \gg 1$. The effects 
of different injection rates  on the decaying part of the light curves are illustrated in the right panel of Fig.~\ref{fig:fig2}. A few things that are worth commenting follow.

\begin{itemize}
\item Fig.~\ref{fig:fig1} (left panel) shows that the SED produced at the peak time of the flare (i.e., $w^{\dpr}\sim \wf$) resembles that of a high-frequency peaked (HSP) blazar \citep{giommiPlanck12} with peak frequency at $\simeq 10^{16}$~Hz.  Spectra obtained during the rise and decay of the flare are similar. Significant differences start to appear at luminosity levels of $2\%$ of the peak bolometric luminosity. The change in the spectral shape during the decay of the flare is caused mainly by adiabatic cooling.  In our scenario, synchrotron (optical--X-rays) and SSC ($\gamma$-ray) flares produced by a single plasmoid are expected to be correlated.    

\item The SED of a flare produced by a monster (large and non-relativistic) plasmoid is shown in Fig.~\ref{fig:fig1} (right panel). Although the characteristics of the particle distribution at injection are the same as in Fig.~\ref{fig:fig1}, the appearance of the SED is different due to electron synchrotron cooling. The magnetic field is the same as in the small plasmoid (left panel in Fig.~\ref{fig:fig1}). Yet,  the synchrotron cooling is  stronger because of the plasmoid's larger size. The effects of cooling are evident already at the peak time of the flare. This explains the lower peak frequency of the low-energy hump ($\sim 10^{14}-10^{15}$~Hz) compared to the left panel of Fig.~\ref{fig:fig1}. The spectra at the rising and decaying part of the flare are different even at times where the luminosity reaches 50\% of the peak bolometric one. In particular, the UV/X-ray and $\gamma$-ray fluxes in the decaying phase decrease rapidly due to the fast cooling of the  radiating electrons at those energy bands, 
which are not being 
replenished. Here, the injection rate of fresh particles  in Phase II is given by eq.~(\ref{eq:Q2}) with $\eta\gg1$. This the most conservative scenario, since it does not require additional free parameters related to the modelling of Phase II. On the contrary, the flux at far-IR wavelengths (or, $\nu\sim 10^{13}$~Hz) decreases more slowly since the radiating  particles have longer cooling timescales. 

\item Fig.~\ref{fig:fig2} (left panel) demonstrates the light curves at two indicative frequency bands that probe the peak ($\nu\sim 10^{16}$~Hz) and the cutoff  ($\sim 10^{17}$~Hz) of the electron synchrotron spectrum. The light curves with the shortest duration and  $\Dtdb=0.22$~hr (solid lines) correspond to a flare produced by a small and relativistic plasmoid, whereas those obtained for a monster plasmoid have  $\Dtdb \sim 2.1$~hr (dashed lines). 
The time ordering of the depicted flares is chosen in such a way as to facilitate the comparison between the two cases.
In both cases, the flux-doubling timescale is in agreement with Fig.~\ref{fig:Dt12_wf} (middle panel).  The rising part of the light curve carries information about Phase I, which is benchmarked with PIC simulations,  whereas the decaying part of the flare is determined by  Phase II.   More realistic, slower decaying flares can also be reproduced by the model for other values of $\eta$ (see right panel in Fig.~\ref{fig:fig2}). We remark that these results are less robust since they dependent on our choice of the magnetic field decay, expansion rate of the 
plasmoid and the functional form for the injection rate in Phase II.   

For the short duration flare produced by the small plasmoid (solid lines), the characteristic timescale of cessation for the particle injection is {chosen to be} $t_{\rm off}\simeq 113$~s.  The photon escape timescale from the blob at the end of Phase I is $t_{\rm cr,f}\sim \wf /c\delta_{\rm p, f} \simeq 190$~s $>t_{\rm off}$. Since the injection of fresh particles shuts off faster than the time it takes for photons produced at the peak of the flare to escape from the plasmoid, the declining part of 
the flare will be dictated by the minimum timescale of $t_{\rm cr, f}$ and the cooling timescale of electrons injected at $\wf$. 

The light curves produced by a bigger and slowly moving plasmoid are presented in Fig.~\ref{fig:fig2} (dashed lines) and are characterized by $\sim$~hr flux-doubling timescale, as predicted in \S\ref{sec:timescale}. Here, $t_{\rm off}\simeq 0.1$~hr and $t_{\rm cr, f}\simeq 0.6$~hr. The decay timescale of the flare, which is defined as the e-folding time of the flux,  is measured to $\simeq 0.1$~hr $< t_{\rm cr, f}$. This suggests that the decay is controlled by the radiative cooling of electrons (see also right panel in Fig.~\ref{fig:fig1}). 

\item The effects of different $\eta$ and $q$ in Phase II on the decaying part of the light curves are illustrated in Fig.~\ref{fig:fig2} (right panel). Different values of $q$ affect the magnetic field strength, whereas $\eta$ controls the cessation timescale as $t_{\rm off}\sim \wf /\eta \vg c \delta_{\rm p, f}$. For $\eta \gg 1$, we obtain the results shown in Figs.~\ref{fig:fig1} and \ref{fig:fig2} (left panel). For a fixed $\eta$ value, we find that the exponent $q$ does not strongly affect the decay timescale, which is mainly determined by adiabatic cooling, but it affects the flux itself; the light curves obtained for $q=2$ lie below those for $q=1$, for all values of $\eta$. The decay timescale decreases as $\eta$ increases, while the dependence on the magnetic field strength weakens; for large $\eta$ the light curves calculated for the two values of $q$ almost coincide. In other words, if no fresh particles are injected soon after the end of Phase 
I,  the decay timescale is set by either the cooling timescale of the last injected particles or the light crossing time of the plasmoid at the end of Phase I. Asymmetric flares with faster decay than rise timescales are expected if the particle injection ceases abruptly.
 \end{itemize}

\begin{figure}
\includegraphics[width=0.48\textwidth]{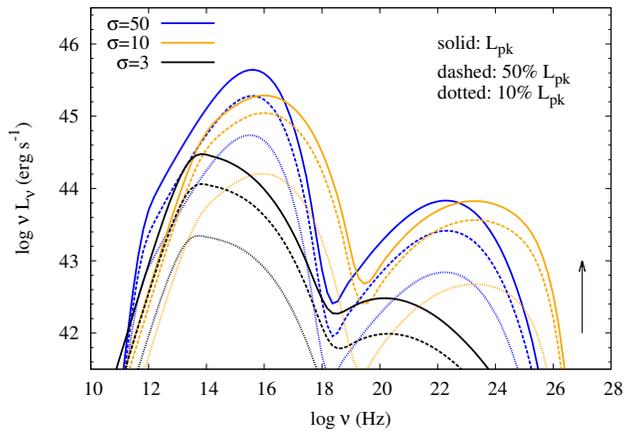}
\caption{Comparison of the multi-wavelength photon spectra obtained during the growth phase of a small and relativistic plasmoid for the three magnetizations considered in this study. The snapshots of the photon spectra correspond to the peak time of the flare (solid lines) and to times where the bolometric luminosity is at 10\% (dotted lines) and 50\% (dashed lines) of its peak value. The arrow shows the time flow. }
\label{fig:fig4}
\end{figure}

\begin{table}
\centering
\caption{Size, momentum and Doppler factor of a small and relativistic plasmoid for $\sigma=3$ and 50. The plasmoid emission is presented in \S\ref{sec:lc-sed}.}
 \begin{tabular}{c c c }
  \hline 
 $\sigma$ & 3 & 50 \\
     \hline
\hline
$w^{\dpr}_0$ (cm) & $8\times 10^{12}$ & $3.5\times 10^{12}$\\
$\wf$ (cm) & $4\times 10^{14}$  & $4\times 10^{14}$\\
$w^{\dpr}_{\rm f}/\ell^\prime$ & {0.04} & 0.04  \\
$\beta_{\rm co}\Gamma_{\rm co}$ & 1.7 & 3.4\\
$\delta_{\rm p, f}$ & 54.6 &  93.1 \\
\hline
 \end{tabular}
 \label{tab3}
\end{table}
A comparison of the multi-wavelength spectra produced by small and relativistic plasmoids in cases of different magnetizations is presented in Fig.~\ref{fig:fig4}.
The properties of the plasmoids, namely their final width, momentum and Doppler factors, are summarized in Table~\ref{tab3}. To make a more direct comparison 
we kept the magnetic field and relativistic electron energy densities equal to their nominal values ($u^{\dpr}_{\rm e}=3\, u_{\rm B}^{\dpr}=0.15$~erg cm$^{-3}$). 
Thus, any changes in the luminosity and spectral shape will be caused by differences in the plasmoid's size, Doppler factor and properties of the particle distribution (i.e. $\gamma_{\min}, \gamma_{\max}$, and $p$). The photon spectra presented in Fig.~\ref{fig:fig4} are snapshots of the flare's emission  during its rising phase.  The spectra calculated for the high-$\sigma$ case are more luminous compared to those for $\sigma=3,10$ mainly due to the higher Doppler factor of the plasmoid (see Tables~\ref{tab2} and \ref{tab3}). The SEDs for $\sigma =10$ and 50 bear some similarities to those of HSP blazars, with the peak frequency being determined by electrons at $\gamma_{\max}$, whereas the synchrotron spectrum for $\sigma=3$ resembles that of LSP blazars. In this case, the luminosity close to the peak of the flare is $\sim 1$ order of magnitude lower  than in the other cases mainly due to the lower $\delta_{\rm p}$.

\section{Discussion}
\label{sec:discussion}
In this paper, we have presented a physically motivated model in the context of relativistic magnetic reconnection  for the emitting regions in blazar jets. We have identified the active region of a blazar jet that gives rise to multi-wavelength flares (from IR/optical wavelengths to TeV $\gamma$-rays) as the site where energy is dissipated through magnetic reconnection. We have argued that the plasmoids that form in the reconnection layer, which may occupy a significant fraction of the jet cross section, are the ``emitting blobs'' of the one-zone leptonic models of blazar emission. We have then presented the fundamental characteristics of the light curves and spectra produced by individual plasmoids. In this section, we discuss various aspects of the model that require further investigation as well as some general remarks on our results.

For the calculation of the plasmoid emission, we have considered synchrotron and SSC radiation. Sources of photons external to individual plasmoids  can, however, greatly enhance the Compton bump of the SED.  The plasmoids that form in a reconnection layer are not isolated but they are members of a plasmoid chain (see e.g.
SGP16). Thus, external Compton (EC) scattering of the radiation produced by other plasmoids becomes relevant. Let us consider two plasmoids: a large and mildly relativistic plasmoid (plasmoid 1) and a small, relativistic one (plasmoid 2) that trails the larger plasmoid.  The effects of EC scattered radiation become important when the relative Lorentz factor $\Gamma_{\rm rel} \gg 1$; this requires that the smaller plasmoid has been substantially accelerated. The synchrotron energy density of the large plasmoid will appear boosted by a factor of $\sim \Gamma^2_{\rm rel}$ in the rest frame of the smaller plasmoid. The energy densities in the respective rest frames are similar. Thus, 
the synchrotron radiation of the larger plasmoid in the rest frame of the smaller one will appear stronger due to the relativistic boosting (for details, see Appendix \ref{app:app1}). In the rest frame of the second plasmoid the total synchrotron energy density is then $\sim (1+\Gamma_{\rm rel}^2) u''_{\rm syn,2}$, where $u''_{\rm syn, 2}$ is the synchrotron energy density produced internally (Appendix \ref{app:app1}).
For $\sigma=10$ and the two plasmoids considered in \S\ref{sec:examples} with sizes $0.04\ell^\prime$ and $0.2\ell^\prime$, we find $\Gamma_{\rm rel}\sim \sqrt{3}$ and the total energy density of synchrotron photons is $\sim 4 u''_{\rm syn,2}$.
Thus, EC scattering may increase the luminosity of the scattered radiation by a factor of several compared to the case where only the internal synchrotron radiation is being up-scattered. For higher magnetizations the relative plasmoid motion can be even more relativistic, thus leading to  higher Compton dominances; in the most extreme case, the energy density of seed photons can increase by a factor $\sim 1 + \sigma$. We thus expect a variety in the Compton dominance of flares powered by plasmoids, especially if the plasmoid statistics are taken into account (e.g., size distribution). We plan to investigate the role of external Compton in the future after including additional photon fields external to the jet.

The particle and magnetic energy densities in the plasmoid are found to be in rough equipartition in PIC simulations that do not self-consistently include radiative cooling (see e.g. SPG15). This is to be compared against our results that include radiative cooling of particles and are obtained assuming macroscopic dimensions of the plasmoids. In general, radiative cooling is more important in plasmoids that reach asymptotically larger sizes (see e.g. left panels in Figs.~\ref{fig:fig1} and Fig.~\ref{fig:fig2}). While $u^{\dpr}_{\rm e} \simeq u^{\dpr}_{\rm B}$ at the time of the plasmoid formation, at later times we find that $u^{\dpr}_{\rm e} \lesssim u^{\dpr}_{\rm B}$ as the particles radiate away their energy. Particle cooling becomes less relevant if the plasmoid Doppler factor is very large $\delta_{\rm p} \gg 10-20$.  This a direct outcome of the relation between the plasmoid size and its momentum and, in this regard, is a distinctive prediction of our model. Given the orientation of 
the reconnection layer (fixed $\theta^\prime$)  and the magnetic field strength in the plasmoid, higher Doppler factors translate to smaller plasmoid sizes, which are characterized, in general, by shorter dynamical timescales. Electron cooling is less significant also for higher magnetizations, where the plasmoids may reach higher asymptotic velocities. This is demonstrated in Fig.~\ref{fig:fig4} where the SED for $\sigma=50$ is compared against those obtained for lower magnetizations.

High Doppler factors ($\gtrsim 50$) are also required to account for the escape of TeV photons \citep[see e.g.][]{finke_08, mastichiadis_08}. This may be problematic for one-zone models, where the emitting blob moves at the jet bulk speed, since $\Gamma_{\rm j}$ would be much larger than the values $10-20$ typically inferred from superluminal motions at pc-scales \citep[e.g.][]{savolainen_10}. However, these extreme Doppler factors are easily obtained in the plasmoid-driven reconnection framework. We showed that for favorable orientations of the reconnection layer and the observer (see e.g. Fig.~\ref{fig:map-ratio}), $\delta_{\rm p}\gtrsim 60$ for small and relativistic plasmoids. In addition, $\delta_{\rm p}$ is larger for higher magnetizations, since $\Gamma_{\rm co}\beta_{\rm co} \rightarrow \sqrt{\sigma}$ (see e.g. eq.~(\ref{eq:PIC}) and Table~\ref{tab2}). We also note that the extreme apparent motions from individual plasmoids are unlikely to be observed because they are expected to slow down to the 
average (bulk) jet motion after exiting the layer. This may resolve the ``Doppler factor crisis'' implied from radio and $\gamma$-ray observations \citep{henri_06}. Furthermore, the radio emission from individual plasmoids is typically suppressed (see e.g. spectra for $\sigma=50$ in Fig.~\ref{fig:fig4}) because of synchrotron self-absorption. 

Radio flares from blazars are often observed to be delayed (on month-long timescales)  compared to powerful $\gamma$-ray flares \citep[e.g.][]{pushkarev_10, hovatta_15, ramakrishnan_15}. A high-energy (X-ray and $\gamma$-ray) short-duration flare in our model, is produced by an individual relatively small plasmoid that forms in the reconnection layer. 
Such plasmoids are unlikely to produce detectable radio flares that coincide with the high-energy flares, since the radio synchrotron emission is typically self-absorbed.
However, delayed radio flares are expected to be produced by all the particles involved in the reconnection process and not only by those that powered the high-energy flare in the first place. In this scenario, the delay timescale between the $\gamma$-ray and the radio flares may be comparable to the overall duration of the reconnection event, which typically amounts to weeks. Studies of reconnection onset and evolution in magnetically dominated jets at large scales are required in order to assess the role of reconnection in powering radio blazar flares.

It is instructive to compare the observed peak luminosity of flares produced by blobs in the traditional one-zone leptonic model (i.e., the blob emission is boosted with $\delta_{\rm j}$) and in the magnetic reconnection framework where the blob may move relativistically in the jet's frame, but grows at a modest fraction ($\beta_{\rm g}\sim0.06-0.1$c) of the speed of light. In the latter, the observed peak luminosity may be written as 
$L_{\rm pl} \sim \dpl^6 (\vg c)^3 t^2 u^{\dpr}$ where $t$ is the observed variability timescale and we neglect  multiplicative factors of order unity. Similarly, for a blob growing at the {\it maximal} possible rate $c$ and being at rest with respect to the bulk jet motion, the observed flare luminosity would be $L \sim \delta_{\rm j}^6 c^3 t^2 u^\prime$. Since $u^\prime \approx u^{\dpr}$ and $\dpl/\delta_{\rm j}\simeq 2\sqrt{\sigma}$ (for a small and fast plasmoid), we find that $L_{\rm pl}/L\sim 64 \, \sigma^4 \vg^3\sim 640 \,  \sigma_1^4 \beta_{\rm g,-1}^3$.  It is clear that for $\sigma \gtrsim 10$ and a given variability timescale $t$ our model may produce far brighter flares than those obtained by one-zone leptonic models.  Large energy densities are often invoked in these models in order to compensate for the lower Doppler boosting.

We have derived analytical expressions for the peak bolometric luminosity and fluence and have shown that they depend on the ratio 
$\ell^\prime/\varpi$. This ratio is also written as $\beta_{\rm rec}/\Gj \theta_{\rm j}$ in the striped-wind scenario for the jet outflow.
Measurements of the jet opening angle at multi-pc scales are available for a substantial sample of
blazars using Very Large Baseline Interferometry (VLBI) imaging. Typically, $\Gj\theta_{\rm j}\simeq 0.1-0.2$ \citep{pushkarev_09, clausen-brown_13}. There is also theoretical evidence that $\Gj \theta_{\rm j} < 1$  for blazar jets. Numerical simulations of acceleration and collimation of
external pressure-supported relativistic jets also find that $\Gj \theta_{\rm j} <1$ after the acceleration is complete \citep{vlahakis_04, komissarov_07, komissarov_09}. Adopting $\theta_{\rm j}\Gj=0.2$, the half-length of the layer is $\ell^\prime/\varpi\sim 0.5$, i.e. the layer occupies a very large fraction of the jet cross section in the striped-wind scenario.

The blazar jet composition is uncertain but several observations \citep[e.g.][]{celotti_08, ghisellini_10, ghisellini_14} are compatible with baryon-loaded jets with several pairs per proton. The multiplicity of pairs $N_{\pm}$ is therefore an additional model parameter that is relevant to reconnection in electron-proton plasmas. $N_{\pm}$ affects the minimum Lorentz factor of the particle distribution for $\sigma \lesssim 10$ (see eq.~(\ref{eq:gmin})), or the maximum one otherwise (\ref{eq:gmax})). So far, we have adjusted the pair-multiplicity in order to obtain synchrotron spectra that are relevant to blazar observations (i.e. peak frequencies in the range $10^{13}-10^{18}$~Hz). In particular, we set $N_{\pm}\sim 1$ for $\sigma\lesssim 10$, whereas for $\sigma=50$ we adopted $N_{\pm}\sim 450$ pais per proton in order to lower $\gamma_{\max}$ down to $\sim 10^3$. Our results suggest an intriguing connection between the magnetization and pair content of blazar jets with the low-energy hump of their SED. 
Given that that there are no PIC simulations similar to those presented in SGP16 for plasmas with $N_{\pm} \gg 1$, it is premature to draw definitive conclusions. For example, it could be possible that the value of $f_{\rm rec}$ in this regime is higher  than the values listed in Table~\ref{tab1}, which, in turn,  would affect $\gamma_{\min}$ and $\gamma_{\max}$. To conclude, the issue of pair-multiplicity in blazar jets in the context of plasmoid-powered emission needs further investigation.

\section{Summary}
\label{sec:summary}
We have presented a physically motivated model for the ``emitting blobs'' in blazar jets in the context of relativistic magnetic reconnection. 
For this purpose, we have combined the results from recent PIC simulations that describe the properties of the plasmoids (e.g. growth rate, acceleration, magnetic field strength and particle number density) with the kinetic equation  for the evolution of the particle distribution and their photon emission. Our approach provides physical insight on basic properties of the plasmoid-powered flares, such as their flux-doubling timescale, while leading to several robust predictions. In particular,  we have shown that correlated synchrotron and SSC flares of duration of several hours--days are powered by the largest and slow-moving plasmoids that form in a reconnection layer. Smaller and fast-moving plasmoids, on the other hand, produce flares of higher peak luminosity, by a factor of $\sim \sigma$, and  of sub-hour duration. Yet, the observed fluence of both types of flares  is similar and depends only weakly on $\sigma$ through the reconnection-related parameters that are well constrained by PIC simulations. 
Multiple flares with a range of flux-doubling timescales (minutes to several hours) observed over a longer period of flaring activity (hours to days) may be used as probes of the layer's orientation and jet's magnetization. 

 \section*{Acknowledgments}
We thank Prof. A. Mastichiadis for providing the numerical code for particle evolution and Dr. S. Dimitrakoudis for producing Figure 1. M.P. is supported by NASA 
through Einstein Postdoctoral Fellowship grant number PF3~140113 awarded by the Chandra X-ray 
Center, which is operated by the Smithsonian Astrophysical Observatory
for NASA under contract NAS8-03060. D.G. acknowledges support from NASA through grant NNX16AB32G issued through the Astrophysics Theory Program. 
We acknowledge access to XSEDE resources under contract No. TG-AST120010, and to NASA High-End Computing 6(HEC) resources through the NASA Advanced Supercomputing (NAS) Division at Ames Research Center. 
\bibliography{blob}

\appendix
\section[]{Solving the electron kinetic equation}
\label{sec:app0}
Here, all calculations  are performed in the plasmoid's rest frame.
To simplify the notation in this section we drop the double primes from all relevant quantities. Let us consider the kinetic equation 
 \eqb
 \label{eq:app0}
 \frac{\partial N}{\partial w} + \frac{\partial}{\partial \gamma} \left(N\frac{{\rm d}\gamma}{{\rm d}w}\right) = Q(\gamma, w)
 \eqe
 that describes the evolution of the electron distribution $N(\gamma, w)\equiv {\rm d}N/{\rm d}\gamma$ under the influence
 of radiative (synchrotron and/or inverse Compton scattering) and adiabatic energy losses, namely
 \eqb
 \frac{{\rm d}\gamma}{{\rm d}w} =  \frac{{\rm d}\gamma}{{\rm d}w}\bigg|_{syn}+\frac{{\rm d}\gamma}{{\rm d}w}\bigg|_{ics}+\frac{{\rm d}\gamma}{{\rm d}w}\bigg|_{ad} < 0.
 \label{eq:app1}
 \eqe
In the absence of a source term in  eq.~(\ref{eq:app0}), 
the number of electrons, $N= \int {\rm d}\gamma N_{\rm G}(\gamma,w)$ is conserved. Here, $N_{\rm G}(\gamma,w)$ 
 is the solution  to eq.~(\ref{eq:app0}) for $Q=0$ and the subscript ``(G)'' is used
 to remind us that it is equivalent to the Green function of the partial differential equation (PDE).
 The number conservation  implies that
 \eqb
 \label{conservation}
 N_{\rm G}(\gamma, w) = N_{\rm G}(\tilde{\gamma}, \tilde{w})\left|\frac{\partial  \tilde{\gamma}}{\partial \gamma}\right|,
 \eqe
 where $ \tilde{\gamma}$ can be written in terms of $\gamma, w$ and $ \tilde{w}$ through the characteristic equation of the PDE (see eq.~(\ref{eq:app1})).
 
 The solution to the non-homogeneous kinetic equation
 can be obtained by convolving the source term with the Green function of the PDE
 \eqb
 N(\gamma, w) = \int^w_0 {\rm d} \tilde{w} Q( \tilde{\gamma}, \tilde{w})\left|\frac{\partial \tilde{\gamma}}{\partial \gamma}\right|,
 \eqe
 where $ \tilde{\gamma}\equiv \tilde{\gamma}(\tilde{w}; \gamma, w)$.
 
 \subsection*{Phase I}
 Adiabatic losses are not important in Phase I and the characteristic equation of the PDE (\ref{kinetic-1}) is therefore written as
 \eqb
 \label{char1}
 \frac{{\rm d}\gamma}{{\rm d}w} = - \ksone \gamma^2
 \eqe
 where  $\ksone$ is defined in eq.~(\ref{ks1}). Here, we consider only synchrotron cooling. The solution of eq.~(\ref{char1}) is then
 \eqb
 \gamma = \frac{ \tilde{\gamma}}{1+ \ksone  \tilde{\gamma} (w- \tilde{w})} 
 \eqe
 or 
 \eqb
 \label{go}
  \tilde{\gamma} = \frac{\gamma}{1-\ksone \gamma (w- \tilde{w})}.
 \eqe
 Combining the above relation with the number conservation eq.~(\ref{conservation}) we find the electron distribution in the absence of 
 a source term 
 \eqb
 N_{\rm G}(\gamma, w)= N_{\rm G}( \tilde{\gamma}, \tilde{w})\left(\frac{ \tilde{\gamma}}{\gamma} \right)^2.
 \eqe
 The solution to the non-homogeneous kinetic equation (\ref{kinetic-1})
 can be obtained by convolving the above relation with the source term of eq.~(\ref{Q1}). The integral to be solved is
 \eqb
 N_I(\gamma, w) = \frac{\pi}{2} n f_p \int_{w_0}^w {\rm d} \tilde{w} \tilde{w}^2\frac{\tilde{\gamma}^{-p+2}}{\gamma^2}H[\tilde{\gamma}-\gmin]H[\gmax-\tilde{\gamma}]
 \eqe
 where $\tilde{\gamma}$ is a function of $\tilde{w}, \gamma, w$.  Substitution of eq.~(\ref{go}) into the above integral results in
 \eqb
  N_I(\gamma, w) = \frac{\pi}{2} n f_p \gamma^{-p} \int_{w_{\rm in}}^{w_{\rm f}} {\rm d}\tilde{w} \tilde{w}^2 \left[1-\ksone \gamma (w-\tilde{w}) \right]^{p-2},
  \label{integral-1}
 \eqe
 where the limits of integration are
 \eqb
 \label{rmin}
 w_{\rm in} & = &  \max\left[w_0, w -\frac{1}{\ksone}\left(\frac{1}{\gamma}-\frac{1}{\gmax} \right) \right] \\
 w_{\rm f} & = &  \min\left[w, w + \frac{1}{\ksone}\left(\frac{1}{\gmin}-\frac{1}{\gamma} \right) \right].
 \label{rmax}
 \eqe
 If the electron does not cool down below $\gmin$, namely $\gamma \ge \gmin$, the upper limit of integration is simply $w_{\rm f}=w$.
 The lower limit of integration is different depending on the electron's Lorentz factor, i.e.
 \eqb
 w_{\rm in} = \left\{ \begin{array}{cc}
                   w_0 & \gamma < \gbr \\
                           &  \\
                    w -\frac{1}{\ksone}\left(\frac{1}{\gamma}-\frac{1}{\gmax} \right)  & \gamma \ge \gbr
                   \end{array}
                 \right.
  \eqe
We note that the cooling break of the electron distribution is derived self-consistently and is given by
\eqb
\label{gbr}
\gbr = \frac{\gmax}{1+\ksone \gmax (w-w_0)}.
\eqe
The calculation of the integral in eq.~(\ref{integral-1}) is then straightforward, and the electron distribution is found to be
\eqb
N_{\rm I}(\gamma, w)= \frac{\pi}{2} n f_p \gamma^{-p} \left[P_1 + P_2 + P_3\right]_{w_{\rm in}}^{w}
\eqe
where 
\eqb
P_1(\tilde{w}; \gamma, w) & = &  \frac{\tilde{w}^2 \left(1-\ksone \gamma (w-\tilde{w}) \right)^{p-1}}{(p-1)\ksone \gamma} \\
P_2(\tilde{w}; \gamma, w) & = & - \frac{2 \tilde{w} \left(1-\ksone \gamma (w-\tilde{w}) \right)^p}{p(p-1)(\ksone \gamma)^2} \\
P_3(\tilde{w}; \gamma, w) & = &  \frac{2 \left(1-\ksone \gamma (w-\tilde{w}) \right)^{p+1}}{(p+1)p(p-1)(\ksone \gamma)^3}
\eqe
or, in a more useful form,
\eqb
\label{N1}
N_{\rm I}(\gamma, w) = \frac{\pi}{2} n f_{\rm p}\gamma^{-p}F_{\rm I}(\gamma,w)
\eqe
where 
\eqb
\label{F1}
F_{\rm I}=G_{\ell}H[\gamma-\gmin]H[\gbr-\gamma]+G_{\rm h}H[\gamma-\gbr]H[\gmax-\gamma].
\eqe

The functions $G_{\ell}$ and $G_{\rm h}$ entering in eq.~(\ref{F1}) are given by 
\eqb
\label{Glow}
G_{\ell} & = & \sum_{i=1}^{3} G_{\ell, i} \\
G_{\ell, 1}& = & \frac{w^2-w_0^2 w_{\ell}^{p-1}}{(p-1)\ksone\gamma} \\ 
G_{\ell, 2} & = & -\frac{2\left(w-w_0 w_{\ell}^p \right)}{p(p-1)(\ksone\gamma)^2}  \\
G_{\ell, 3} & = & \frac{2\left(1- w_{\ell}^{p+1}\right)}{(p+1)p(p-1)(\ksone\gamma)^3},
\eqe
and 
\eqb
\label{Ghigh} 
G_{\rm h} & = & \sum_{i=1}^{3} G_{\rm h, i} \\
G_{\rm h, 1} & = & \frac{w^2 -\left(\frac{\gamma}{\gmax}\right)^{p-1}w_{\rm h}^2}{(p-1)\ksone\gamma}  \\
G_{\rm h, 2} & = &  - \frac{2w-2\left(\frac{\gamma}{\gmax}\right)^{p} w_{\rm h}}{p(p-1)(\ksone\gamma)^2} \\ \nonumber 
G_{\rm h, 3} & = &  \frac{2-2\left(\frac{\gamma}{\gmax}\right)^{p+1}}{(p+1)p(p-1)(\ksone\gamma)^3}.
\eqe
Finally, the functions $w_{\ell}$ and $w_{\rm h}$ are defined by
\eqb
w_{\ell}(\gamma, w) = 1- \ksone\gamma(w-w_0).
\eqe
and
\eqb
w_{\rm h}(\gamma, w) = w- \frac{1}{\ksone\gamma}\left(1-\frac{\gamma}{\gmax} \right).
\eqe 

\subsection*{Phase II}
Adiabatic and synchrotron energy losses are relevant in this phase. The characteristic equation for the particle's Lorentz factor
is given by eq.~(\ref{char2}) and its solution for $2q +\alpha \neq 0$ is
 written as
 \eqb
 \label{char-ad-syn}
 \gamma =\tilde{\gamma} \left(\frac{\tilde{w}}{w}\right)
 \left[1+ \frac{\kstwo \tilde{\gamma} \tilde{w}^{1+s}}{s}\left(\left(\frac{w}{\tilde{w}}\right)^{s}-1 \right)\right]^{-1}
 \eqe
 where $s=-2q-\alpha$. The effect of combined synchrotron and adiabatic losses 
 enters through the term enclosed by the parenthesis. As expected, for $\kstwo=0$ we retrieve 
 the evolution of $\gamma$ in the presence of adiabatic losses only.
  The electron distribution in the absence of a source term evolves as 
 \eqb
 \frac{N_{\rm G}(\gamma, w)}{N_{\rm G}(\tilde{\gamma}, \tilde{w}) } = 
 \left( \frac{w}{\tilde{w}}\right)\left[1+ \frac{\kstwo \tilde{\gamma} \tilde{w}^{1+s}}{s}
 \left( \left(\frac{w}{\tilde{w}}\right)^{s}  -1 \right)\right]^2.
 \label{homo2}
 \eqe
  Note that the effect of synchrotron losses appears only in the second term of the r.h.s. of eq.~(\ref{homo2}). 
 By setting $\kstwo=0$ we obtain the solution  to the pure adiabatic loss case. 
The solution to the non-homogeneous kinetic equation (\ref{kinetic-2})
 can be obtained by convolving the source term (see eq.~(\ref{Q2})) with the Green function of the PDE, i.e.
 \eqb
 \label{full2}
 N_{\rm II}(\gamma, w) = \int^w \!\!\! {\rm d}\tilde{w} Q_{\rm II}(\tilde{\gamma}, \tilde{w})\left( \frac{w}{\tilde{w}}\right)\left[1+ \frac{\kstwo \tilde{\gamma} \tilde{w}^{1+s}}{s} \left( \left(\frac{w}{\tilde{w}}\right)^{s}  -1 \right)\right]^2,
 \eqe
 where $\tilde{\gamma}\equiv\tilde{\gamma}(\tilde{w}; \gamma, w)$. 
 
Using eq.~(\ref{N1}) and eq.~(\ref{full2}) we derive the electron distribution in Phase II:
 \eqb
\label{N2}
 N_{\rm II}(\gamma, w) = \frac{\pi n f_{\rm p}}{2 w_{\rm f}} \left(\frac{w}{w_{\rm f}} \right)^{-(p-1)}\gamma^{-p} F_{\rm e}^{p-2} F_{\rm II}
 \eqe
 where 
  \eqb
 \label{Fe}
 F_{\rm e}(\gamma, w)=1-\frac{\kstwo \gamma w_{\rm f}^{1+s}}{s}\left(\frac{w}{w_{\rm f}}\right)\left(\left(\frac{w}{w_{\rm f}}\right)^{s}-1 \right),
 \eqe
with $F_{\rm II}$ defined as
 \eqb
 \label{F2}
 F_{\rm II} & = &  G_{\ell}(\gstar, w_{\rm f})H[\gstar-\gmin]H[\gbr(w_{\rm f})-\gstar] + \\ \nonumber 
 & & G_{\rm h}(\gstar, w_{\rm f})H[\gstar-\gbr(w_{\rm f})]H[\gmax-\gstar].
 \eqe
In the above expression $\gstar$ is defined as 
 \eqb
 \gstar \equiv \frac{\gamma}{F_{\rm e}(\gamma, w)}\frac{w}{w_{\rm f}}.
 \eqe
 The Heavyside functions that appear in eq.~(\ref{F2}) will determine the time-evolution
 of the lower and upper cutoffs of the electron distribution, as well as that of the cooling break. In brief, we find
 \eqb
 \label{gmax2}
 \tilde{\gamma}_{\max}(w) = \gmax\left(\frac{w_{\rm f}}{w} \right)
 \left[1+ \frac{\kstwo \gmax w_{\rm f}^{1+s}}{s}
 \left( \left(\frac{w}{w_{\rm f}}\right)^{s}  -1 \right) \right]^{-1},
 \eqe
 and similar for $\tilde{\gamma}_{\min}(w)$. The cooling break energy evolves
 as 
 \eqb
 \tilde{\gamma}_{\rm br}(w)=\gbr(w_{\rm f})\left(\frac{w_{\rm f}}{w} \right)
 \left[1+ \frac{\kstwo \gbr(w_{\rm f}) w_{\rm f}^{1+s}}{s}
 \left( \left(\frac{w}{w_{\rm f}}\right)^{s}  -1 \right) \right]^{-1},
 \eqe
where $\gbr$ is defined in eq.~(\ref{gbr}).

\section[]{A naive estimate of the flux-doubling timescale}
\label{app:naive}
\begin{figure}
\centering
\includegraphics[width=0.48\textwidth]{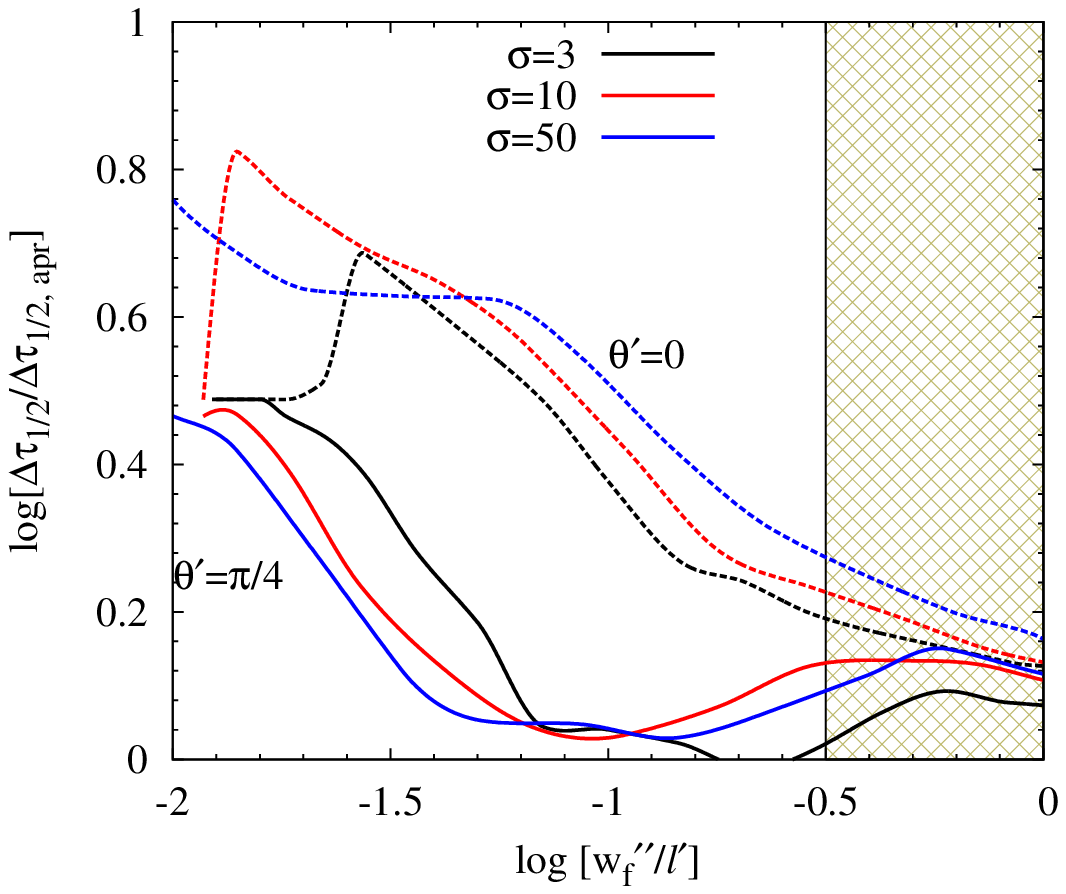}
 \caption{Plot of the ratio $\Delta \tau_{1/2}/\Delta \tau_{1/2,\rm apr}$ (in logarithmic units)  as a function of $\wf/\ell^\prime$ for the three magnetizations considered in this study and two values of $\theta^\prime$, i.e. $\pi/4$ (solid lines) and 0 (dashed lines). Coloured curves correspond to different magnetizations (see inset legend). Here, $\thobs=0.5/\Gj$. }
\label{fig:comparison}
\end{figure}

A naive estimate of the flux-doubling timescale, $\Delta \tau_{1/2,\rm apr}$, can be obtained if the acceleration of the plasmoid and the suppression of its growth rate are ignored (see eq.~(\ref{eq:estimate})). 
The ratio $\Delta \tau_{1/2}/\Delta \tau_{1/2,\rm apr}$ (in logarithmic units) is presented in Fig.~\ref{fig:comparison} for the three magnetizations considered in this study and two values of $\theta^\prime$, i.e. $\pi/4$ (solid lines) and 0 (dashed lines). 
We find that the expression (\ref{eq:estimate}) underestimates (by at least a factor of three) the actual doubling timescale for flares produced by plasmoids with small sizes ($\wf \ll w_{\rm f,c}^{\dpr}$). For favorable orientations, e.g. $\thobs=0.5/\Gj$ and $\theta^\prime=\pi/4$ (solid lines) the differences are caused mainly by the suppression of the growth rate, which is not included in eq.~(\ref{eq:estimate}). The reason is that plasmoids that leave the layer with such small sizes have been already accelerated significantly and, thus growing at a lower rate than $\vg$  (by a factor of 2-3 for the particular choice of the suppression factor $g(X^\prime/w^{\dpr})$). For non-favorable orientations, e.g. $\thobs=0.5/\Gj$ and $\theta^\prime=0$, where small changes in the plasmoid momentum have larger impact on its Doppler factor, the differences between $\Dtdb$ and $\Delta \tau_{\rm 1/2, apr}$ are larger. The two expressions tend to become similar for plasmoids with large sizes ($\wf > w_{\rm f,c}$) and non-
relativistic motions. 

\section[]{Effect of relative plasmoid motion on the synchrotron energy densities}
\label{app:app1}
Let us consider two plasmoids: a large and mildly relativistic plasmoid (plasmoid 1)  and a small, relativistic one (plasmoid 2) that trails the larger plasmoid. In this paragraph we adopt the following notation: subscripts  1 and 2 refer to quantities of the first and second plasmoids, respectively. Single and double primes denote quantities measured in their respective comoving frames.

The energy density of synchrotron photons in the rest frame of the two plasmoids is given by
\eqb
u^\prime_{\rm syn, 1} = \frac{\nu_{\rm p} L_{\rm syn}(\nu_{\rm p}) \large|_{1} }{4\pi cR_{\rm I, 1}^2 \delta_{\rm p,1}^4}, \quad u^{\dpr}_{\rm syn, 2}= \frac{\nu_{\rm p} L_{\rm syn}(\nu_{\rm p})\large|_{2} }{4\pi c R_{\rm I,2}^2 \delta_{\rm p,2}^4},
\label{eq1}
\eqe
where we approximated the total synchrotron luminosity by its value at the peak frequency $\nu_{\rm p}$ in each case. {
To exemplify the effect of external Compton scattering on the spectra shown in \S\ref{sec:examples} in a qualitative manner,  henceforth we adopt $0.2\ell^\prime$ and $0.04\ell^\prime$ as the sizes of the large and small plasmoids, respectively. }By inspection of the SEDs in Figs.~\ref{fig:fig1} and \ref{fig:fig2} we estimate that $\nu_{\rm p} L_{\rm syn}(\nu_{\rm p}) \large|_{1} / \nu_{\rm p} L_{\rm syn}(\nu_{\rm p}) \large|_{2} \sim 0.5$. Using also the values in Table~\ref{tab2} for the sizes and Doppler factors we find
that 
\eqb
\frac{u'_{\rm syn, 1}}{u''_{\rm syn, 2}} \sim 0.5
\eqe
One can come to a similar conclusion by noting that the electron energy density is independent of the plasmoid size ($u^\prime_{\rm e, 1}\simeq u^{\dpr}_{\rm e, 2}$)
and that is radiated mostly as synchrotron radiation (e.g., $u'_{\rm syn, 1}\sim u^\prime_{\rm e, 1}$).  
To assess the role of EC scattering of the synchrotron radiation from plasmoid 1, we calculate the quantity $u^{\dpr}_{\rm syn,1}$. 
Let us suppose that the velocity vectors of the two components are parallel. Then their relative velocity and Lorentz factor
are given by 
\eqb
\beta_{\rm rel}=\frac{\beta_{\rm co,2}-\beta_{\rm co,1}}{1-\beta_{\rm co,1} \beta_{\rm co,2}}, \quad \Gamma_{\rm rel} = \Gamma_{\rm co,1} \Gamma_{\rm co,2} \left(1-\beta_{\rm co,1} \beta_{\rm co,2} \right).
\eqe
For example, using the values for $\sigma=10$, i.e. $\Gamma_{\rm co,1}\simeq 1.2$ and $\Gamma_{\rm co, 2} =2.6$ (see also Table~\ref{tab2})  we find
that the relative motion is mildly relativistic $\beta_{\rm rel}=0.8$ and $\Gamma_{\rm rel}=1.7$. For higher $\sigma$-plasmas higher $\Gamma_{\rm rel}$ can be achieved.  

Using the invariance of $u(\epsilon, \mu)/\epsilon^3$ and the 
transformation of the solid angle 
\eqb
{\rm d}\Omega''& = & \frac{2\pi}{\Gamma_{\rm rel}^{2} \left(1-\beta_{\rm rel} \mu'\right)^{2}} {\rm d}\mu' 
\eqe
we find that 
\eqb
u''_{\rm syn,1} & = & \int {\rm d}\epsilon''\!\! \int d\mu'' u''_{\rm syn,1}(\epsilon'', \mu'') = \nonumber \\
& = & \int {\rm d}\epsilon' \int {\rm d}\mu' \Gamma_{\rm rel}^2\left(1-\beta_{\rm rel} \mu'\right)^2 u'_{\rm syn,1}(\epsilon', \mu') = \nonumber \\
& = & \frac{\Gamma_{\rm rel}^2 u'_{\rm syn,1}}{2}\int_{-1}^{\mu{12}}{\rm d}\mu' \left(1-\beta_{\rm rel} \mu'\right)^2
\eqe
where an isotropic synchrotron photon field in the comoving frame of plasmoid 1 was assumed, i.e. $u'_{\rm syn,1}(\mu')=u'_{\rm syn,1}/2$, and
$\mu_{12} = X_{12} / \sqrt{X_{12}^2+R_{\rm I, 1}^2}$; here, $X_{12}$ is the separation distance of the plasmoids.
The integral is function of $X_{12}$ and $\beta_{\rm rel}$ given by
\eqb
I(X_{12}, \beta_{\rm rel}) = 1-\mu_{12} + \beta_{\rm rel}\left(1+\mu^2_{12}\right)+\frac{\beta_{\rm rel}^2}{3}\left(1+\mu_{12}^3\right).
\eqe
For $X_{12} \ll R_{\rm I,1}$, $I(X_{12}, \beta_{\rm rel}) \rightarrow 1+\beta_{\rm rel}+(1/3)\beta_{\rm rel}^2 \sim 2$. Similarly, if the second plasmoid has approached the leading one at a distance comparable to its radius ($\mu_{12}=1/\sqrt{2}$) and $I \sim 2$. Thus, $u''_{\rm syn,1} \simeq \Gamma_{\rm rel}^2 u'_{\rm syn,1}\approx \Gamma_{\rm rel}^2 u''_{\rm syn,2}$. In the rest frame of plasmoid 2 the total synchrotron energy density is then $u''_{\rm tot,2}\simeq (1+\Gamma_{\rm rel}^2) u''_{\rm syn,2} \sim 4 u''_{\rm syn,2}$. Thus, EC scattering may increase the luminosity of the scattered radiation by a factor of $\sim 4$ compared to the case where only the internal synchrotron radiation is being up-scattered.
{We note that even larger enhancements of the Compton bump of the SED are expected for higher magnetizations and/or different plasmoid sizes.}

\end{document}